\documentclass[10pt,a4paper]{article}
\usepackage{graphicx,amsmath,amsfonts,amssymb}

\usepackage[T2A]{fontenc}

\renewcommand{\Re}{\mbox{Re}}

\begin{document}

\vspace*{3cm}

\begin{center}

{\large \bf COULOMB CORRECTIONS TO THE PARAMETERS OF THE
LANDAU---POMERANCHUCK---MIGDAL EFFECT THEORY AND ITS ANALOGUE}

\smallskip

{\it O. Voskresenskaya, E. Kuraev, H. Torosyan}

\smallskip

{\small Joint Institute for Nuclear Research, Joliot--Curie 6,
141980 Dubna, Moscow Region, Russia}

\end{center}

{\small Using the Coulomb correction to the screening angular
parameter of the Moli\`{e}re multiple scattering theory, we obtained
analytically and numerically the Coulomb corrections to the
parameters of the Migdal LPM effect theory. We showed that these
corrections allow to eliminate the discrepancy between the
predicti\-ons of the LPM effect theory and its  measurement at least
for high Z targets and also to further improve the agreement between
the predicti\-ons of the LPM effect theory analogue for a thin layer
of matter and experimental
data.\\


\vspace{.5cm}

\section*{Introduction}

Landau and Pomeranchuk were the first to show \cite{C-2} that
multiplicity of electron scattering processes by atomic nuclei in an
amorphous medium results in the suppression of soft bremsstrahlung.
The quantitative theory of this phenomenon was created by Migdal
\cite{C-4,C-4.5}\footnote{See also \cite{C-5} accounting the edge
effects.}. Therefore, it received the name
Landau--Pomeranchuk--Migdal  (LPM) effect.

The next step in the development of the quantitative theory of the
LPM effect was made in \cite{Operat1} on the basis of the
quasi-classical operator method in QCD \cite{Operat2}. One of the
basic equations of this method is the Schr\"{o}dinger equation in
the external field with an imaginary potential, which admits of
formal solution in the form of the path integral. The path integral
treatment of the LPM effect was proposed and developed in [7--12].

It was shown that analogous effects are possible also at coherent
radiation of relativistic electrons and positrons in a crystalline
medium \cite{Sh1}, in cosmic-ray physics  \cite{Klein1} (e.g. in
applications motivated by extremely high energy IceCubes
neutrino-induced showers with energies above 1 PeV \cite{Klein2}).
Effects of this kind should manifest themselves in scattering of
protons by the nuclei, which has recently been shown in Groning by
the AGOR collaboration \cite{C-1.2}, as well as at penetration of
quarks and partons through the nuclear matter \cite{KST}. The QCD
analogue of the LPM effect was examined in \cite{Z1,Gyu,QCD}; a
possibility studying the LPM effect in oriented crystal at GeV
energy was analyzed in \cite{Kryst}; theoretically, an analogue of
the LPM effect was considered for nucleon-nucleon collisions in the
neutron stars, supernovae \cite{stars}, and relativistic plasmas
\cite{Plasma}.

The results of a series of experiments at the SLAC
\cite{C-1.1.1,C-1.1.2,Klein3} and CERN-SPS  \cite{C-1.3,C-1.4}
accelerators on detection of the Landau--Pomeranchuk effect
confirmed the basic qualitative conclusion that multiple scattering
of ultrarelativistic charged particles in matter leads to
suppression of their bremsstrah\-lung in the soft part of the
spectrum. However, attempts to quantitatively describe the
experimental data \cite{C-1.1.1} faced an unexpec\-ted difficulty.
For achieving satisfactory agreement of data with theory
\cite{C-4,C-4.5}, the authors \cite{C-1.1.1} had to multiply the
 results of their calculations in the Born approximation by the
 normalization factor $R$ equal to $0.94\pm
0.01\pm 0.032$, which had no reasonable explanation.

The alternate calculations \cite{C-3,Cont1} gave a similar result
despite different computational basis \cite{C-1.1.1}. The
theoretical predictions are in agreement with the spectrum of photon
brems\-strahlung measured for 25 GeV electron beam and
$0.7-6.0\%L_{\scriptscriptstyle
\mathrm{R}}$\footnote{$L_{\scriptscriptstyle \mathrm{R}}$ presents a
radiation length of a target material here.} gold target over the
range $30 <\omega< 500$ MeV of the emitted photon frequency $\omega$
only within a normalization factor $0.93$ \cite{C-3} -- $0.94$
\cite{C-1.1.1}. The origin of the above small but significant
disagreement between data and theory needs to be better understood
\cite{C-1.1.2}. In  \cite{Z2} the further develop\-ment of the
light-cone path integral approach to PLM effect was performed. The
Coulomb effects as well as multiphoton emission and absorbtion was
taken into account. A detailed comparison with SLAC E-146 data was
carried out. Nevertheless, the problem of normali\-zation remained
and is still not clear. The other authors,  except \cite{C-3,Z2}, do
not discuss normalization \cite{Klein3}.

The aim of this work is to show that the discussed discrepancy
between data and theory can be explained at least for high Z targets
if the corrections to the results of the Born approximation (i.e.,
the Coulomb corrections) are appropriately considered on the basis
of a revised version of the Moli\`{e}re multiple scattering theory
\cite{C-11.1,C-11.2}. The paper is organized as follows. In Section
1 we consider the basic formulae of the quantitative LPM effect
theory for finite-size targets obtained by the kinetic equation
method and also the small-angle approximation of this theory which
is used further for analytical and numerical calculations. In
Section 2 we present the results of the conventional \cite{C-7} and
a revised Moli\`{e}re multiple scattering theory
\cite{C-11.1,C-11.2} applied  in the next Section to the theory of
the LPM effect and its analogue for a thin target
\cite{Shul'ga,Fomin}. In Section 3 we obtain the analytical and
numerical results for Coulomb corrections to the quantities of the
LPM effect theory and its analogue for a thin layer of matter in
some asymptotic cases and also in the regimes corresponding to the
conditions of the experiment \cite{C-11.1,C-11.2}. Finally, we
summarize our findings and state our conclusions.

\section{\Large{LPM effect theory  for finite
targets}}

There exist two methods that allow one to develop a rigo\-rous
quantitative theory of the Landau--Pomeranchuk effect. This is
Migdal's method of kinetic equation \cite{C-4,C-4.5}  and the method
of functional integration [7--12,31]. Neglecting numerically small
quantum-mechanical corrections, we will adhere to version of the
Landau--Pomeranchuk effect theory, developed in \cite{C-4, C-5,
Sasha}.

\subsection{\normalsize \textbf{Basic formulae}}

Simple though quite cumbersome calculations using the results
\cite{C-4,C-5} yield the following formula for the electron spectral
bremsstrahlung intensity averaged over various trajectori\-es of
electron motion in an amorphous medium (hereafter the units  $\hbar
= c =1$, $e^2=1/137$ are used) \cite{Sasha}:
\begin{equation*}\left\langle\frac{dI}{d\omega}\right\rangle =
2\sum_{\boldsymbol\epsilon} \biggl\{n_0^{} L \int f^{\ast}(\mathbf{
n}_2)\nu(\mathbf{n}_2-\mathbf{n}_1)f(\mathbf{n}_1) d\mathbf{n}_1 d
\mathbf{n}_2\end{equation*}
\begin{equation*}-(n_0^{}v)^2\int\limits_{0}^{T}dt_1\int\limits_{t_1}^{T}dt_2 \,\,
\Re\biggl [ \int f^{\ast}(\mathbf{n}_2)\nu(\mathbf{n}_2-
\mathbf{n}_2^{\prime})f(\mathbf{n}_1)\end{equation*}
\begin{equation}\label{1} \times
\nu(\mathbf{n}_1^{\prime}-\mathbf{n}_1)w(t_2,t_1,\mathbf{n}_2^{\prime},
\mathbf{n}_1^{\prime},\mathbf{k}) d\mathbf{n}_1d\mathbf{n}_1^{\prime}d\mathbf{n}_{2}d
\mathbf{n}_{2}^{\prime}\biggl ]\biggl\},
\end{equation}
where
\begin{equation*}f(\mathbf{n}_{1,2})=\frac{e}{2\pi}\cdot\frac {\boldsymbol\epsilon
\mathbf{v}_{1,2}}{1-\mathbf{n} \cdot \mathbf{v}_{1,2}}\ ,
\end{equation*}
\begin{equation*}\mathbf{v}_{1,2}=v\cdot \mathbf{n}_{1,2}\ ,\quad\mathbf{n} =
\frac{\mathbf{k}}{\omega}\ ,\quad d\mathbf{n}_{1,2} \equiv do_{1,2}\ ,\quad
T=\frac{L}{v}\ ,
\end{equation*}
\begin{equation*}\nu(\mathbf{n}_2-\mathbf{n}_1)=\delta(\mathbf{n}_2-\mathbf{n}_1)
\int \sigma_0(\mathbf{n}_2^{\prime}-\mathbf{n}_1)d\mathbf{n}_2^{\prime}
-\sigma_0(\mathbf{n}_2-\mathbf{n}_1)\ ,
\end{equation*}
\begin{equation*}
w(t_2,t_1,\mathbf{n}_2,\mathbf{n}_1,\mathbf{k})= \int \tilde
 w(t_2,t_1,\mathbf{r}_2- \mathbf{r}_1, \mathbf{n}_2, \mathbf{n}_1)
%\end{equation*}
%
%\begin{equation*}\\
%\times
\exp\left[i\omega(t_2-t_1)-i\mathbf{k}(\mathbf{r}_2-\mathbf{r}_1)\right]d\mathbf{r}_2\
 .\end{equation*}
Here $\boldsymbol\epsilon$ and $\mathbf{k}$ are the polarization
vector and the wave vector of the emitted photon; $n_0$ denotes the
density of the scattering centers per unit length of fast scattered
particle trajectory, $L $ is the target thickness,
$\mathbf{n}_{1,2}$ are the unit vectors in the electron motion
direction, $\mathbf{v}$ and $v$ are the electron velocity assumed to
be invariant during the interaction with the target (the
quantum-mechanical recoil effect is negligibly small) and its
modulus, $e$ is the electron charge, $\sigma _0(\mathbf{n}_2 -
\mathbf{n}_1)\, =\, d\sigma /do_{\mathbf{n}_2}$ presents the
differential Born cross section of the electron scattering by target
atoms. The direction of motion $\mathbf{n}_2$ at time $t_2$ provided
that at the time $t_1$ the electron had the coordinate
$\mathbf{r}_1$ and moved in the direction characterized by the unit
vector $\mathbf{n}_1$. The electron distribution function in the
coordinate $ \mathbf{r_2}$, $w(t_2,t_1,\mathbf{r}_2-\mathbf{r}_1,
\mathbf{n}_2, \mathbf{n}_1)$, satisfies the kinetic equation
\begin{equation*}\frac{\partial w(t_2,t_1,\mathbf{r}_2-\mathbf{r}_1,\mathbf{n}_2,
\mathbf{n}_1)}{\partial t_2}=-\mathbf{v}_2 \cdot
\boldsymbol\nabla_{\mathbf{r}_2}\cdot
w(t_2,t_1,\mathbf{r}_2-\mathbf{r}_1,\mathbf{n}_2, \mathbf{n}_1)
\end{equation*}
\begin{equation}\label{2} -
n_0\int\nu(\mathbf{n}_2-\mathbf{n}_2^{\prime}) \tilde
w(t_2,t_1,\mathbf{r}_2-
\mathbf{r}_1,\mathbf{n}_2^{\prime},\mathbf{n}_1)d\mathbf{n}_2^{\prime}
\end{equation}
with the boundary condition
\begin{equation}\label{3} \tilde w(t_2,t_1,\mathbf{r}_2-\mathbf{r}_1,
\mathbf{n}_2,\mathbf{n}_1) \vert
_{t_2=t_1} = \delta (\mathbf{r}_2 -\mathbf{r}_1)\delta (\mathbf{n}_2 -\mathbf{n}_1).
\end{equation}

The term of (\ref{1}) linear in  $n_0$ is a `usual' (incoherent)
contribution to the intensity of the electron bremsstrahlung in the
medium, derived by summation of the radiation intensities of the
electron interaction with separate atoms of the target. The term
quadratic in   $n_0$ includes  the contribution from the
interference of the bremsstrahlung amplitudes on various atoms. The
destructive character of this interference leads to suppression of
the soft radiation intensity, i.e. to the Landau--Pomeranchuk
effect.

For $\omega$ larger than
$\omega_{cr}=4\pi\gamma^2/(e^2L_{\scriptscriptstyle \mathrm{R}})$,
where $\gamma$ is the Lorentz factor of the scattered particle and
$L_{\scriptscriptstyle \mathrm{R}}$ is the radiation length of the
target material (for estimation of $\omega_{cr}$, see
\cite{C-2,C-4,Z2,Shul'ga}\footnote{In the conditions of experiment
\cite{C-1.1.1,C-1.1.2}, $\omega_{cr}\approx 244$ MeV for
$0.7-6.0\%L_{\scriptscriptstyle \mathrm{R}}$ gold target at 25 GeV
(see Table 1 in \cite{Z2}).}), the interference term becomes
negligibly small, and radiation is of pure incoherent character.

\subsection{\normalsize \textbf{Small-angle approximation}}

For ultra-relativistic particles $(1-v\ll 1)$ it is convenient to
pass in (\ref{1}) to the small-angle approximation
($\vartheta_{1,2}\ll 1$) according to the scheme

\begin{equation*}\mathbf{n}_{1,2}=\biggl(1-\frac{\vartheta_{1,2}^{2}}{2}\biggl )\mathbf{n} +
\boldsymbol{\vartheta}_{1,2}, \quad
d\mathbf{n}_{1,2}=d\boldsymbol{\vartheta}_{1,2}\ ;\end{equation*}
\begin{equation} f(\mathbf{n}_{1,2})=f(\boldsymbol{\vartheta_{1,2}})=\frac{e}{\pi}\cdot\frac {
\boldsymbol{\epsilon}\,\boldsymbol{\vartheta}_{1,2}}{\vartheta_{1,2}^{2}+\lambda^2},\quad
\lambda = \frac{m}{E} = \gamma^{-1}\ ;
\end{equation}
\begin{equation*}\sigma_{0}(\mathbf{n}_2-\mathbf{n}_1)=\sigma_{0}(\boldsymbol{\vartheta}_2-
\boldsymbol{\vartheta}_1),\quad
\delta(\mathbf{n}_2-\mathbf{n}_1)=\delta(\boldsymbol{\vartheta}_2-
\boldsymbol{\vartheta}_1)\ ,
\end{equation*}
\begin{equation*}\nu(\mathbf{n}_2-\mathbf{n}_1)=
\nu(\boldsymbol{\vartheta}_2-\boldsymbol{\vartheta}_1), \quad
\boldsymbol{\vartheta}_2-\boldsymbol{\vartheta}_1=\boldsymbol{\theta}\ ;
\end{equation*}
\begin{equation*} w(t_2,t_1,\mathbf{n}_2,\mathbf{n}_1,\mathbf{k}) =
w(t_2,t_1,\boldsymbol{\vartheta}_2, \boldsymbol{\vartheta}_1,\omega)
\end{equation*}
\noindent and further to the Fourier transforms of $f,\,\nu \, , w$
%:
\begin{equation*}f(\boldsymbol{\eta}) =  \frac{1}{2\pi}\int \tilde f(\boldsymbol{\theta})
\exp[i\boldsymbol{\eta}\boldsymbol{\theta}]d\boldsymbol{\theta} = \frac{ie\lambda
\boldsymbol{\epsilon}\,\boldsymbol{\eta}} {\pi\eta}K_{1}(\lambda\eta)\ ;
\end{equation*}
\begin{equation}\label{5}
\nu(\eta) =  \int \tilde
 \nu(\boldsymbol{\theta})e^{i\boldsymbol{\eta}\boldsymbol{\theta}}d\boldsymbol{\theta}
=
2\pi\int\sigma_0(\boldsymbol{\theta})[1-J_0(\eta\theta)]\boldsymbol{\theta}
d\boldsymbol{\theta};
\end{equation}
\begin{equation*}w(t_2,t_1,\boldsymbol{\eta}_2,\boldsymbol{\eta}_1,\omega) =
\frac{1}{(2\pi)^2}\int \tilde
 w(t_2,t_1,\boldsymbol{\vartheta}_2,\boldsymbol{\vartheta}_1,\omega)
\end{equation*}
\begin{equation*}
\times \exp[i\boldsymbol{\eta}_2\boldsymbol{\vartheta}_2 -
i\boldsymbol{\eta}_1\boldsymbol{\vartheta}_1
]d\boldsymbol{\vartheta}_1 d\boldsymbol{\vartheta}_2\
,\end{equation*}\\
where $\boldsymbol{\vartheta}_{1(2)}$ denotes a two-dimensional
electron scattering angle in the plane orthogonal to the electron
direction at instant of time $t_{1(2)}$, $m$ and $E$ are the
electron mass and its energy, $\boldsymbol{\theta}$ presents the
electron multiple scattering angle over the time interval $t_2-t_1$,
$\lambda$ is the characteristic frequency of the emitted photon,
$J_0$ and $K_1$ are the Bessel and Macdonald functions,
respectively.

Consequently, expression (\ref{1}) is reduced to

\begin{eqnarray}\left\langle\frac{dI}{d\omega}\right\rangle &=& \frac{2\lambda^2 e^2}{\pi^2}
\biggl\{ n^{}_0 L \int K^{2}_{1}(\lambda\eta)\nu(\eta)
d\boldsymbol{\eta}\nonumber\\
&&-n_0^2\int\limits_{0}^{L}dt_1\int\limits_{0}^{L}dt_2 \int
\frac{(\boldsymbol{\eta}_1\boldsymbol{\eta}_2 )}{\eta_1 \eta_2}
K_{1}(\lambda\eta_1)K_{1}(\lambda\eta_2)\nu(\eta_1)\nu(\eta_2)\nonumber\\
\label{6} &&\times\Re
[w(t_2,t_1,\boldsymbol{\eta}_2,\boldsymbol{\eta}_1,\omega)]
d\boldsymbol{\eta}_1d\boldsymbol{\eta}_{2}\biggl\}\ ,\end{eqnarray}\\
where $w$ satisfies the kinetic equation
\begin{eqnarray}\frac{\partial w(t_2,t_1,\boldsymbol{\eta}_2,
\boldsymbol{\eta}_1,\omega)}{\partial t_2}
&=&\frac{i\omega}{2}\,(\lambda^2 -
\Delta_{\boldsymbol{\eta}_2})w(t_2,t_1,\boldsymbol{\eta}_2,\boldsymbol{\eta}_1,\omega)\nonumber\\[2mm]
\label{7} &&- n^{}_0\nu(\eta_2)w(t_2,t_1,\boldsymbol{\eta}_2,
\boldsymbol{\eta}_1,\omega)
\end{eqnarray}\\
or, equivalently,
\begin{eqnarray}i\,\frac{\partial w(t_2,t_1,\boldsymbol{\eta}_2,
\boldsymbol{\eta}_1,\omega)}{\partial t_2}& =&
\left[\frac{\omega}{2}
\Delta_{\boldsymbol{\eta}_2}-\frac{\omega}{2}\lambda^2-in_0^{}\nu(\eta_2)\right]\nonumber\\[2mm]
\label{7.5} &&\times
w(t_2,t_1,\boldsymbol{\eta}_2,\boldsymbol{\eta}_1,\omega)
\end{eqnarray}\\
with the boundary condition
\begin{equation}
\label{8} w(t_2,t_1,\boldsymbol{\eta}_2,\boldsymbol{\eta}_1,\omega)
= \delta(\boldsymbol{\eta}_2 -\boldsymbol{\eta}_1).
\end{equation}\\
The form of (\ref{7.5}) is similar to the equation for Green's
function of the two-dimensional Schr\"{o}dinger equation with the
mass $\omega^{-1}$ and the complex potential
\begin{equation}
\label{9} U(\eta) = - \frac{\omega\lambda^2}{2} - i\,n^{}_0
\nu(\eta)
\end{equation}\\
and therefore admits of a formal solution in the form of a continual
integral (see, e.g., \cite{C-6}). The analysis of (\ref{6}) will be
continued in Section 3.

\section{\Large\textbf{Multiple scattering theory}}

The theory of the multiple scattering of charged particles has been
treated by several authors. However, most widespread at present is
the multiple scattering theory of Moli\`{e}re \cite{C-7,Bethe}. The
results of this theory are employed nowadays in most of the
transport codes. It is of interest for numerous applications related
to particle transport in matter, and it also presents the most used
tool for taking into account the multiple scattering effects in
experimental data processing.

As the Moli\`{e}re theory is currently used roughly for $10-300$ GeV
electron beams, the role of the high-energy corrections to the
parameters of this theory becomes significant. Of special importance
is the Coulomb correction to the screening angular parameter,  as
this parameter also enters into other important quantities of the
Moli\`{e}re theory.

\subsection{\normalsize \textbf{Moli\`{e}re's theory of multiple scattering}}

Let $w_{\scriptscriptstyle \mathrm{M}}(\vartheta,L)$ be a
spatial-angle particle distribution function in a homogenous medium,
and $\boldsymbol{\vartheta}$ is a two-dimensional particle
scattering angle in the plane orthogonal to the incident particle
direction. For small-angle approximation $\vartheta\ll 1$
($\sin\vartheta \sim \vartheta$), the above distribution function is
the number of particles scattered in the angular interval
$d\vartheta$ after traveling through the target of thickness $L$. In
the notation of Moli\`{e}re, it reads
\begin{equation}\label{14}
w_{\scriptscriptstyle \mathrm{M}}(\vartheta,L)= \int\limits_0^\infty
J_0(\vartheta \eta)\exp[-n^{}_{0}L\cdot\nu(\eta)] \eta\, d\eta\ ,
\end{equation}
where
\begin{equation}
\label{15} \nu(\eta)=2\pi \int \limits_0^\infty
\sigma_0(\boldsymbol{\theta})[1-J_0(\theta \eta)]\boldsymbol{\theta}
d\boldsymbol{\theta}\ .
\end{equation}

The function (\ref{14}) satisfies the well-known  Boltzmann
transport equation, written here with the small-angle approximation
\begin{eqnarray}\label{Boltzmann}
\frac{\partial w(\vartheta,L)}{\partial
L}&=&-n_0\,w_{\scriptscriptstyle
\mathrm{M}}(\vartheta,L)\int\sigma_0(\boldsymbol{\theta})d^2\boldsymbol{\theta}
+ n_0\int w_{\scriptscriptstyle
\mathrm{M}}(\boldsymbol{\vartheta}+\boldsymbol{\theta},L)\sigma_0(\boldsymbol{\theta})
d^2\boldsymbol{\theta}\nonumber\\
&=&n_0\int \left[w_{\scriptscriptstyle
\mathrm{M}}(\boldsymbol{\vartheta}+\boldsymbol{\theta},L)-w_{\scriptscriptstyle
\mathrm{M}}(\vartheta,L)\right]\sigma_0(\boldsymbol{\theta})
d^2\boldsymbol{\theta} \ .
\end{eqnarray}
The Gaussian particle distribution function used in the Migdal LPM
effect theory, which differs from (\ref{14}), can be derived from
the Boltzmann transport equation by the method of Fokker and Planck
\cite{C-9}.

One of the most important results of the Moli\`{e}re theory is that
the scattering is described by a single parameter, the so-called
screening angle ($\theta_a$ or $\theta_a^{\,\prime}$)
\begin{equation}\theta_a^{\,\prime}=\sqrt{1.167}\,\theta_a=
\left[\exp\left(C_{\scriptscriptstyle
\mathrm{E}}-0.5\right)\right]\theta_a\approx1.080\,\theta_a\ ,
\end{equation}
where $C_{\scriptscriptstyle \mathrm{E}}=0.577\ldots$ ~is the Euler
constant.

More precisely, the angular distribution depends only on the
logarithmic ratio $b$,
\begin{equation}\label{b} b=\ln \left(\frac{\theta_c}{\theta_a^{\,\prime}} \right)^2\equiv\ln
\left(\frac{\theta_c}{\theta_a} \right)^2+1-2C_{\scriptscriptstyle
E}\ , \end{equation}
of the characteristic angle $\theta_c$ describing the foil thickness
\begin{equation}\label{char} \theta_c^2=4\pi n^{}_0L\left(\frac{Z\alpha}{\beta p}
\right)^2,\quad p=mv\ , \end{equation}
to the screening angle $\theta_a^{\,\prime}$, which characterizes
the scattering atom.

In order to obtain a result valid for large angles,  Moli\`{e}re
defines a new parameter $B$ by the transcendental equation
\begin{equation}\label{B} B-\ln B=b\ . \end{equation}
The angular distribution function can then be written as
\begin{equation*} w_{\scriptscriptstyle \mathrm{M}}(\vartheta,B) =\frac{1}{
\overline{\vartheta^2}}\int\limits_0^{\infty}y dy J_0 (\vartheta
y)e^{-y^2/4}\end{equation*}
\begin{equation}\label{exp2}
\times\exp\left[\frac{y^2}{4B}\ln\left(\frac{y^2}{4}\right)\right],\quad
y=\theta_c\eta\ .
\end{equation}

The Moli\`{e}re  expansion method is to consider the term
$y^2\ln(y^2/4)/4B$ as a small parameter.  Then, the angular
distribution function  is expanded in a power series in $1/B$:
\begin{equation}\label{power} w_{\scriptscriptstyle
\mathrm{M}}(\vartheta,L)=\sum\limits_{n=0}^{\infty}\frac{1}{n!}\frac{1}{B^n}w_n(\vartheta,L)\
,
\end{equation}
in which
\begin{equation} w_n(\vartheta,L) =
\frac{1}{\overline{\vartheta^{\,2}}}\int\limits_0^{\infty}y dy J_0
\left(\frac{\vartheta}{\sqrt{\overline{\vartheta^2}}}\, y\right)
e^{-y^2/4}\left[\frac{y^2}{4}\ln\left(\frac{y^2}{4}\right)\right]^n\
,
\end{equation}
\begin{equation}\label{vartheta2}\overline{\vartheta^{\,2}}=\theta_c^2B=4\pi
n^{}_0L\left(\frac{Z\alpha}{\beta p} \right)^2B(L)\ .
\end{equation}
This method is valid for $B\geq 4.5$ and
$\overline{\vartheta^{\,2}}<1$.

The first function $w^{}_0(\vartheta,L)$ has a simple analytical
form
\begin{equation}\label{W_0} w^{}_0(\vartheta,L)=\frac{2}{\overline{
\vartheta^{\,2}}}\exp\left(\!\!-\frac{\vartheta^2}{\overline{\vartheta^{\,2}}}\right),
\end{equation}
\begin{equation}\overline{\vartheta^{\,2}} \mathop{\sim}\limits_{\;\;L\,\to \,\infty}\;
\frac{L}{L_{\scriptscriptstyle \mathrm{R}}}\ln \left(
\frac{L}{L_{\scriptscriptstyle \mathrm{R}}} \right).
\end{equation}
For small angles, i.e., $\vartheta/\overline{\vartheta}=
\vartheta/(\theta_c\sqrt{B})$ less than about 2, the Gaussian
\eqref{W_0} is the dominant term.  In this region,
$w_1(\vartheta,L)$ is in general less than $w_0(\vartheta,L)$, so
that the correction to the Gaussian is of order of $1/B$, i.e.,
about $10\%$.

A good approximate representation of the distribution at any angle
is
\begin{equation}\label{2order} w_{\scriptscriptstyle \mathrm{M}}(\vartheta,L) =w_0(\vartheta,L)+
\frac{1}{B}w_1(\vartheta,L)\end{equation}
with
\begin{equation}\label{W_1} w_1(\vartheta,L)=
\frac{1}{\overline{\vartheta^{\,2}}}\int\limits_0^{\infty}y dy J_0
\left(\frac{\vartheta}{\sqrt{\overline{\vartheta^2}}}\, y\right)
e^{-y^2/4}\left[\frac{y^2}{4}\ln\left(\frac{y^2}{4}\right)\right]\ .
\end{equation}
This approximation was applied  by authors of  \cite{Fomin} to the
analysis of data \cite{C-1.1.1,C-1.1.2} over the region $\omega<30$
MeV that will be shown in Section 3.

Let us notice that the expression (\ref{15}) for the function
$\nu(\eta)$ is identical to (\ref{5}). As was shown in classical
works of Moli\`{e}re \cite{C-7}, this quantity can be represented in
the area of the important $\eta$ values  $0\leq \eta\leq 1/\theta_c$
as
\begin{equation}\label{16} \nu(\eta)=-4\pi\Bigg(\frac{Z\alpha}{\beta p}\Bigg)^2
\eta^2\,\left[
\ln\left(\frac{\eta\,\theta_a}{2}\right)+C_{\scriptscriptstyle
\mathrm{E}}-\frac{1}{2}\right]\ ,
\end{equation}
where the screening angle $\theta_a$ depends both on the screening
properties of the atom and on the $\sigma_0(\boldsymbol{\theta})$
approximation used for its calculation.

Using the Thomas--Fermi model of  the atom and an interpolation
scheme, Moli\`{e}re obtained $\theta_a$ for the cases where
$\sigma_0(\boldsymbol{\theta})$ is calculated within the Born and
quasi-classical approximations:
\begin{equation}\label{17}
\theta_a^{\scriptscriptstyle \mathrm{B}}=1.20\cdot\alpha\cdot
Z^{1/3}\ ,
\end{equation}
\begin{equation}\label{18}
\theta_a^{\scriptscriptstyle
\mathrm{M}}=\theta_a^{\scriptscriptstyle \mathrm{B}}\sqrt{1+3.34\,
(Z\alpha/\beta)^2}\ .
\end{equation}
The latter result is only approximate (see critical remarks on its
derivation in \cite{C-9}). Below we will present an exact analytical
and numerical result for this angular parameter.

\subsection{\normalsize\textbf{Coulomb correction to the screening
angular parameter}}

Recently, it has been shown \cite{C-11.2} by means of \cite{Operat1}
that for any model of the atom the following rigorous relation
determining  the screening angular parameter $\theta^\prime_a$ is
valid:
\begin{equation*}
\ln(\theta^\prime_a)=\ln(\theta^\prime_a)^{\scriptscriptstyle
\mathrm{B}} + \Re\left[ \psi
(1+iZ\alpha/\beta)\right]+C_{\scriptscriptstyle \mathrm{E}}
\end{equation*}
or, equivalently,
\begin{equation}\label{basres}
\Delta_{\scriptscriptstyle
\mathrm{CC}}[\ln\big(\theta_a^{\,\prime}\big)]\equiv
\ln(\theta^\prime_a)-\ln(\theta^\prime_a)^{\scriptscriptstyle
\mathrm{B}} =f(Z\alpha/\beta)\ ,
\end{equation}
where $\Delta_{\scriptscriptstyle \mathrm{CC}}$ is the so-called
Coulomb correction to the Born result, $\psi$ is the logarithmic
derivative of the gamma function $\Gamma$, and $f(Z\alpha/\beta)$ is
an universal function of the Born parameter $\xi=Z\alpha/\beta$
which is also known as the Bethe--Maximon function:
\begin{equation}\label{summa} f(\xi)=\xi^2\sum_{n=1}^\infty\frac{1}{n(n^2+\xi^2)}\ .
\end{equation}

To compare the approximate  Moli\`{e}re result (\ref{18}) for the
Coulomb correction with the exact one (\ref{basres}), we first
present (\ref{18}) in the form
\begin{eqnarray}
\label{del2}\delta^{}_{\scriptscriptstyle \mathrm{
M}}[\theta_a]\equiv\frac{\theta^{\scriptscriptstyle
\mathrm{M}}_a-\theta_a^{\scriptscriptstyle \mathrm{B}}}
{\theta_a^{\scriptscriptstyle \mathrm{B}}}  =\sqrt{1+3.34\, \xi^2}-1
\end{eqnarray}
and also rewrite (\ref{basres}) as follows:
\begin{eqnarray}\label{del0}
\delta_{\scriptscriptstyle
\mathrm{CC}}[\theta_a]\equiv\frac{\theta_a-
\theta_a^{\scriptscriptstyle \mathrm{B}}}
{\theta_a^{\scriptscriptstyle
\mathrm{B}}}=\frac{\theta_a^{\,\prime}-
\big(\theta_a^{\,\prime}\big)^{\scriptscriptstyle \mathrm{B}}}
{\big(\theta_a^{\,\prime}\big)^{\scriptscriptstyle \mathrm{B}}}=
\exp\left[f\left(\xi\right)\right]-1\ .
\end{eqnarray}
Then we get
\begin{eqnarray}\label{corr3}\Delta_{\scriptscriptstyle
\mathrm{CCM}}[\delta]&\equiv& \delta^{}_{\scriptscriptstyle
\mathrm{CC}}[\theta_a]-\delta^{}_{\scriptscriptstyle
\mathrm{M}}[\theta_a]\ , \\
\label{corr3-1} \delta^{}_{\scriptscriptstyle
\mathrm{CCM}}[\delta]&\equiv&\frac{\Delta_{\scriptscriptstyle
\mathrm{CCM}}[\delta]}{\delta^{}_{\scriptscriptstyle
\mathrm{M}}[\theta_a]} \,.
\end{eqnarray}

In order to obtain relative difference between the  approximate
$\theta_a^{\scriptscriptstyle \mathrm{M}}$ and exact $\theta_a$
results for the screening angle
\begin{eqnarray}\label{angle}
\delta_{\scriptscriptstyle
\mathrm{CCM}}[\theta_a]&\equiv&\frac{\theta_a-
\theta_a^{\scriptscriptstyle \mathrm{M}}}
{\theta_a^{\scriptscriptstyle \mathrm{M}}}=\frac{\theta_a}
{\theta_a^{\scriptscriptstyle \mathrm{M}}}-1\\
\label{angle-1} &=& R_{\scriptscriptstyle \mathrm{CCM}}[\theta_a]-1\
,
\end{eqnarray}
we rewrite definitions (\ref{del2}), (\ref{del0}) in the following
form
\begin{equation}
\label{folform}
\delta_{\scriptscriptstyle\mathrm{CC}}[\theta_a]+1=\frac{\theta_a}
{\theta_a^{\scriptscriptstyle\mathrm{B}}}\ ,\qquad
\delta^{}_{\scriptscriptstyle \mathrm{
M}}[\theta_a]+1=\frac{\theta^{\scriptscriptstyle\mathrm{M}}_a}
{\theta_a^{\scriptscriptstyle\mathrm{B}}}\end{equation}
and obtain for the ratio $R_{\scriptscriptstyle
\mathrm{CCM}}[\theta_a]$ the expression
\begin{eqnarray}
\label{rat} R_{\scriptscriptstyle
\mathrm{CCM}}[\theta_a]&\equiv&\frac{\theta_a}
{\theta_a^{\scriptscriptstyle
\mathrm{M}}}=\frac{\delta_{\scriptscriptstyle\mathrm{CC}}
[\theta_a]+1}{\delta^{}_{\scriptscriptstyle \mathrm{
M}}[\theta_a]+1}
\end{eqnarray}
\begin{equation}
=\delta_{\scriptscriptstyle \mathrm{CCM}}[\delta]+1\ .
\end{equation}
We can also represent the relative difference (\ref{angle}) by the
equation
\begin{equation}\label{reldif}
\delta_{\scriptscriptstyle
\mathrm{CCM}}[\theta_a]=\frac{\Delta_{\scriptscriptstyle
\mathrm{CCM}}[\delta]}{\delta^{}_{\scriptscriptstyle \mathrm{
M}}[\theta_a]+1}\ .
\end{equation}

For some high Z targets used in \cite{C-1.1.2} and $\beta=1$, we
obtain the following values of the relative Moli\`{e}re
$\delta_{\scriptscriptstyle \mathrm{M}}[\theta_a]$ (\ref{del2}) and
Coulomb $\delta_{\scriptscriptstyle \mathrm{CC}}[\theta_a]$
(\ref{del0}) corrections and also the sizes of the relative
differences $\delta_{\scriptscriptstyle \mathrm{CCM}}[\delta]$
(\ref{corr3-1}), $\delta_{\scriptscriptstyle
\mathrm{CCM}}[\theta_a]$ (\ref{angle-1} and the
ratio $R_{\scriptscriptstyle \mathrm{CCM}}[\theta_a]$ (\ref{rat}) (Table 1).

\begin{center} {\bf Table 1.}
Numerical results for the relative corrections (\ref{del2}),
(\ref{del0}), relative differences (\ref{corr3-1}), (\ref{angle-1}),
and the ratio (\ref{rat}) in the range of nuclear charge $73\leq
 Z \leq 92$.
\end{center}

\begin{center}
\begin{tabular}{ccccccc}
\hline \\[-3mm]
Target&Z&$\delta_{\scriptscriptstyle
\mathrm{M}}[\theta_a]$&$\delta_{\scriptscriptstyle
\mathrm{CC}}[\theta_a]$& $\delta_{\scriptscriptstyle
\mathrm{CCM}}[\delta]$&$10\,\delta_{\scriptscriptstyle
\mathrm{CCM}}[\theta_a]$&$R_{\scriptscriptstyle
\mathrm{CCM}}[\theta_a]$ \\[.2cm]
\hline\\[-3mm]
Ta&73&0.396&0.318&$-0.198$&$-0.562$&0.944\\
W &74&0.404&0.325&$-0.196$&$-0.565$&0.943\\
Pt&78&0.443&0.359&$-0.189$&$-0.582$&0.942\\
Au&79&0.452&0.367&$-0.188$&$-0.585$&0.941\\
Pb&82&0.482&0.393&$-0.185$&$-0.600$&0.940\\
U& 92&0.583&0.485&$-0.169$&$-0.622$&0.938\\[.2cm]
\hline
\end{tabular}
\end{center}

From the Table 1 it is evident that the Coulomb correction
$\delta_{\scriptscriptstyle \mathrm{CC}}[\theta_a]$ has a large
value, which ranges from around $30\%$ for $Z\sim 70$ up to $50\%$
for $Z\sim 90$. The relative difference between the approximate and
exact results for this Coulomb correction varies from 17 up to
$20\%$ over the range $73\leq Z\leq 92$. The relative difference
$\delta_{\scriptscriptstyle CCM}[\theta_a]$ between the approximate
$\theta^{\scriptscriptstyle\mathrm{M}}_a$ and exact $\theta_a$
results for the screening angle as well $R_{\scriptscriptstyle
CCM}[\theta_a]=\theta_a/\theta^{\scriptscriptstyle\mathrm{M}}_a$
value does not vary significantly from one target material to
another. Their sizes are $5.86\pm 0.22\%$ for
$-\delta_{\scriptscriptstyle \mathrm{CCM}}[\theta_a]$ and $0.941\pm
0.002$ for $R_{\scriptscriptstyle \mathrm{CCM}}[\theta_a]$ in the Z
range studied.

It is interesting that the latter value coincides with the
normalization constant $R=0.94\pm 0.01$ found in \cite{C-1.1.1}. We
show further that the above discrepancy between theory and
experiment \cite{C-3,C-1.1.1,C-1.1.2} can be eliminated on the basis
of these Coulomb corrections to the screening angular parameter at
least for heavy target elements.

\section{\Large\textbf{Coulomb corrections
in the LPM effect theory and its analogue for a thin layer of
matter}}

\vspace{.5cm}

\subsection{\normalsize\textbf{Coulomb corrections
to  the parameters of the LPM effect theory for finite targets}}

Analytical solving (\ref{7}) with arbitrary values of $\omega$ is
only possible within the Fokker--Planck approximation\footnote{An
explicit expression for $w$ obtained in this approach can be found
in \cite{C-5}.}
\begin{equation}
\label{10}\nu(\eta) = a\cdot \eta^{2},
\end{equation}
at  $\omega = 0$ it is also  possible for arbitrary  $\nu (\eta)$.

In the latter case  $(\omega = 0)$
\begin{equation}\label{11}
w(t^{}_2,t^{}_1,\boldsymbol{\eta}^{}_2,\boldsymbol{\eta}^{}_1,0) =
\delta (\boldsymbol{\eta}^{}_2 -\boldsymbol{\eta}^{}_1)
\exp[-n^{}_0\nu(\eta^{}_2)(t^{}_2-t^{}_1)]~,
\end{equation}\\
and integration over $t^{}_1,\,t^{}_2$ in (\ref{6}) is carried out
trivially, leading to the simple result\\
\begin{equation}\label{12}
\left\langle\frac{dI}{d\omega}\right\rangle\biggl\vert_{\omega=0} =
\frac{4\lambda^2 e^2}{\pi} \int K^{2}_{1}(\lambda\eta)
\left\{1-\exp[{-n^{}_0}L\nu(\eta)]\right\}\eta d\eta\ .
\end{equation}\\

Considering the aforesaid, in the other limiting case $(\omega \gg
\omega_{cr})$ we get
%:
\begin{equation}\label{13}
\left\langle\frac{dI}{d\omega}\right\rangle\biggl\vert_{\omega \gg
\omega_{cr}} = n^{}_0 L\lambda^2 e^2 \int K^{2}_{1}
(\lambda\eta)\nu(\eta)\eta d\eta \ .
\end{equation}

\subsubsection{\normalsize{Case $\boldsymbol\omega
\mathbf{\gg}\boldsymbol\omega_{cr}$}}

After the substitution of $\nu(\eta)$ (\ref{16}) into (\ref{13}),
the integration is carried out analytically, leading to the
following result:
\begin{equation}\label{22}
\left\langle\frac{dI}{d\omega}\right\rangle\biggl\vert_{\omega \gg
\omega_{cr}} = \frac{16}{3\pi}\cdot \frac{Z^2\alpha^3}{m^2} \cdot
\biggl(\ln\frac{\lambda}{\theta_a}+ \frac{7}{12}\biggl )\cdot
\,n_0\, L\ .
\end{equation}\\

Let us find an analytical expression for the Coulomb correction to
the Born spectral bremsstrah\-lung rate (\ref{22}):
\begin{equation*}
\Delta_{\scriptscriptstyle \mathbf{CC}}\big[\left\langle
dI/d\omega\right\rangle\big]\equiv\left\langle\frac{dI}{d\omega}\right\rangle-
\left\langle\frac{dI}{d\omega }\right\rangle^{\scriptscriptstyle
\mathrm{B}}
\end{equation*}\\
\begin{equation}\label{22.5}
=-\frac{16\,Z^2\alpha^3\,n_0 L}{3\pi\, m^2}\cdot\biggl [\ln
(\theta^\prime_a)-\ln(\theta^\prime_a)^{\scriptscriptstyle
\mathrm{B}}\biggl] = -\frac{16\,Z^2\alpha^3\, n_0 L}{3\pi\, m^2}
\cdot f(\xi)\ .
\end{equation}\\

Then the corresponding relative Coulomb correction reads
\begin{equation*}\label{22.7}
\delta_{\scriptscriptstyle \mathrm{CC}}\big[\left\langle
dI/d\omega\right\rangle\big]\equiv \frac{\left\langle
dI/d\omega\right\rangle- \left\langle dI/d\omega
\right\rangle^{\scriptscriptstyle
 \mathrm{B}}}{\left\langle dI/d\omega \right\rangle^{\scriptscriptstyle
\mathrm{B}}}
\end{equation*}
\begin{equation}\label{22.7}
= -\frac{f(\xi)}{0.583-\ln\left(1.2\,\alpha\,Z^{1/3}\right)}\ .
\end{equation}\\

Let us enter the ratio
\begin{equation}\label{23}
R_{\scriptscriptstyle \mathrm{CC}}(\omega)=\frac{\left\langle
dI(\omega)/d\omega \right\rangle} {\,\,\left\langle
dI(\omega)/d\omega \right\rangle^{\scriptscriptstyle \mathrm{B}}} =
\delta_{\scriptscriptstyle \mathrm{CC}}\big[\!\left\langle
dI/d\omega\right\rangle\!\big]+1\ .
\end{equation}\\
We will now estimate the numerical values of (\ref{22.7}) and
(\ref{23}) (Table 2).

It will seen from Table 2 that the relative correction to the Born
spectral bremsstrahlung rate is about $-8\%$. Whereas the
calculations of Blancenbeckler and Drell \cite{Cont1} reproduce the
Migdal results for thick  targets with the $+8\%$ higher emission
probability when the interference term vanishes. Therefore it is
natural to normalize these calculations by means of the obtained
Coulomb correction $\bar \delta_{\scriptscriptstyle
\mathrm{CC}}\big[\left\langle dI/d\omega\right\rangle \big]=-7.97\pm
0.71\%$. The corresponding ratio $R(\omega)\vert_{\omega \gg
\omega_{cr}}$ is approximately 0.92 for the gold target\footnote{The
use of approximate Moli\`{e}re's result (\ref{18}) or (\ref{basres})
for $\theta_a$ would give the value $R(\omega)\vert_{\omega
\gg\omega_{cr}}=0.900 $ in the discussed case.}  discussed in
\cite{C-1.1.1}.

\newpage

\begin{center} {\bf Table 2.}
The relative Coulomb correction $\delta_{\scriptscriptstyle
\mathrm{CC}}\big[\left\langle dI/d\omega\right\rangle \big]$ to the
Born spectral bremsstrahlung rate for some high Z targets,
$\omega\gg\omega_{cr}$, and $\beta=1$.
\end{center}

\begin{center}

\begin{tabular}{lrcccc}
\hline \\[-3mm]
Target\!\!&\!\!Z&$Z\alpha$&$f(Z\alpha)$&$-\delta_{\scriptscriptstyle
\mathrm{CC}}$& $R_{\scriptscriptstyle \mathrm{CC}}
$\\[.2cm]
\hline\\[-3mm]
~~W&74&0.540&0.281&$0.072$&0.928\\
~~Au&79&0.577&0.313&$0.081$&0.919\\
~~Pb&82&0.598&0.332&$0.086$&0.914\\[.2cm]
\hline
\end{tabular}

\medskip

$\bar \delta_{\scriptscriptstyle \mathrm{CC}}\big[\left\langle
dI/d\omega\right\rangle \big]=-7.97\pm 0.71\%$

\end{center}

This value coincides within the $3.2\%$ systematic error with the
value of normalization factor $R=0.94\pm 0.1\pm 0.32$, which was obtained in
\cite{C-1.1.1} for the $0.7-6\%L_{\scriptscriptstyle \mathrm{R}}$
gold target in the region $450<\omega<500$ MeV\footnote{Migdal used
a Gaussian approximation for multiple scattering. This
underestimates the probability of large angle scatters. These
occasional large angle scatters would produce some suppression for
$\omega>\omega_{cr}$, where Migdal predicts no suppression and where
the authors of \cite{C-1.1.1} determine the normalization
\cite{C-1.1.2}.}.

\subsection{\normalsize{Case $\boldsymbol\omega\mathbf{=0}$}}

In the other limiting case the performance of numerical integration
in (\ref{12}) get the following results for the relative Coulomb
correction $-\delta_{\scriptscriptstyle
\mathrm{CC}}\big[\left\langle dI/d\omega\right\rangle\big]$ and the
ratio $R(\omega)\vert_{\omega=0}$ (Table 3) at thicknesses
$L=0.7-6\%L_{\scriptscriptstyle \mathrm{R}}$ of experimental gold
targets \cite{C-1.1.1}.

\medskip

\begin{center} {\bf Table 3.}
The relative correction $\delta_{\scriptscriptstyle
\mathrm{CC}}\big[\left\langle dI/d\omega\right\rangle\big]$
 for $Z=79$ and $\omega=0$.
\end{center}

\begin{center}
\begin{tabular}{ccc}
\hline \\[-3mm]
~~~L(cm)&$-\delta_{\scriptscriptstyle \mathrm{CC}}\big[\left\langle
dI/d\omega\right\rangle\big] $&$R_{\scriptscriptstyle
 \mathrm{CC}}\big[\left\langle dI/d\omega\right\rangle\big]$\\[.2cm]
\hline\\[-3mm]
~~~$0.060\,L_{\scriptscriptstyle
\mathrm{R}}$&~~~~~0.018&0.982\\
~~~$0.007\,L_{\scriptscriptstyle
\mathrm{R}}$&~~~~~0.039&0.961\\[.2cm]
\hline
\end{tabular}
\end{center}

\medskip

\noindent Here $L_{\scriptscriptstyle \mathrm{R}}\approx 0.33$ cm is
the radiation length of the target material $(Z=79)$
\begin{equation}
L_{\scriptscriptstyle
\mathrm{R}}=\frac{4Z^2e^6n^{}_0}{m^2}\ln\left(183Z^{1/3}\right)\ .
\end{equation}

\subsection{\normalsize{Case $
\boldsymbol\omega_{cr}\mathbf{>}\boldsymbol\omega $}}

When $\omega_{cr}> \omega > 0$, it is obvious from general
considerations that
\begin{equation}\label{25}
R_{\scriptscriptstyle \mathrm{CC}}(\omega)\vert_{\omega >
\omega_{cr}}\leq R_{\scriptscriptstyle
\mathrm{CC}}(\omega)\vert_{\omega_{cr}>\omega} \leq
R_{\scriptscriptstyle \mathrm{CC}}(\omega)\vert_{\omega=0}\ .
\end{equation}
From Table 3 and (\ref{25}) it follows that the calculation results
for $\left\langle dI/d\omega\right\rangle$ cannot be obtained from
the Born approximation results by multiplying them by the
normalization constant, which is independent of the frequency
$\omega$ and target thickness $L$.

However, considering a nearly $3.2\% $ systematic error of the
experimental data \cite{C-1.1.1} in the range $500 >\omega>30$ MeV,
it is clear why multiplication by the normalization factor helped
the authors of \cite{C-3,C-1.1.1} to get reasonable agreement of the
Born calculation results with the experimental data.

In the conditions of the experiment \cite{C-1.1.1,C-1.1.2,Klein3},
it is permissible to draw conclusions about the size of the
normalization factor based on the corrections to the Bethe--Heitler
spectrum in the frequency range approximately from 244 to 500 MeV
(for 25 GeV beam and $0.7\% L_{\scriptscriptstyle \mathrm{R}}$ gold
target). It is, although some caution is advisable, since 244 to 500
MeV is a rather narrow range. Therefore, let us consider also the
second limiting case in order to obtain some interpolation values
for $R_{\scriptscriptstyle
\mathrm{CC}}(\omega)\vert_{\omega_{cr}>\omega}$ from Tables 2 and 3
(Table 4).

\newpage

\begin{center} {\bf Table 4.}
The interpolation values of the ratio $R_{\scriptscriptstyle
\mathrm{CC}}(\omega,L)$ for $\omega<\omega_{cr}$, $Z=79$ (Au), and
$\beta=1$.
\end{center}

\begin{center}
\begin{tabular}{lcc}
\hline \\[-3mm]
~~~L(cm)&~~~~$R_{\scriptscriptstyle \mathrm{CC}}\vert_{\omega >
\omega_{cr}}\leq R_{\scriptscriptstyle
\mathrm{CC}}\vert_{\omega_{cr}>\omega} \leq R_{\scriptscriptstyle
\mathrm{CC}}\vert_{\omega=0}$ &~~~~$ \bar R_{\scriptscriptstyle
\mathrm{CC}}(\omega)\vert_{\omega
< \omega_{cr}}$\\[.2cm]
\hline\\[-3mm]
~~~$0.007\,L_{\scriptscriptstyle \mathrm{R}}$&~~~~$0.920\leq
R_{\scriptscriptstyle \mathrm{CC}}(\omega,L)\vert_{\omega
< \omega_{cr}} \leq 0.961$&0.940~~~~\\
~~~$0.060\,L_{\scriptscriptstyle \mathrm{R}}$&~~~~$0.920\leq
R_{\scriptscriptstyle \mathrm{CC}}(\omega,L)\vert_{\omega
< \omega_{cr}} \leq 0.982$&0.951~~~~\\[.2cm]
\hline\\[-3mm]
\end{tabular}
\smallskip

$\bar R_{\scriptscriptstyle \mathrm{CC}}(\omega,L)\vert_{\omega <
\omega_{cr}}=0.945\pm0.08~~~~~$

\end{center}

So for  $0.007\,L_{\scriptscriptstyle \mathrm{R}}$ to
$0.060\,L_{\scriptscriptstyle \mathrm{R}}$ gold targets, the mean
value of the ratio $R_{\scriptscriptstyle
\mathrm{CC}}(\omega,L)\vert_{\omega < \omega_{cr}}$ is approximately
$0.945\pm 0.008$, which coincides within the experimental error with
the normalization factor value $0.94\pm 0.01\pm 0.032$ introduced in
\cite{C-1.1.1} for obtaining agreement of the calculations performed
in the Born approximation with experiment. The obtained result means
that the normalization is not required for the spectral density of
radiation $\left\langle dI(\omega)/d\omega\right\rangle$ calculated
on the basis of the refined screening angle.

We will now obtain the analytical expressions and numerical
estimations for the Coulomb corrections to the function
$\nu(\eta)=2\pi\int\sigma_0(\boldsymbol{\theta})[1-J_0(\eta\theta)]\boldsymbol{\theta}
d\boldsymbol{\theta}$ (\ref{5}) and the complex potential $U(\eta)=-
\omega\lambda^2/2 - i\,n^{}_0 \nu(\eta)$ (\ref{10}).

For the first quantity, using (\ref{16}), we have
\begin{equation*}
\Delta_{\scriptscriptstyle \mathrm{CC}} [\nu(\eta)]\equiv
\nu(\eta\,)- \nu^{\scriptscriptstyle
\mathrm{B}}(\eta\,)\end{equation*}
\begin{equation}\label{nu}=-4\pi \eta^{\,2} \left(Z\alpha/\beta
p\right)^2\Delta_{\scriptscriptstyle
CC}[\ln\big(\theta_a^{\,\prime}\big)] =-4\pi \eta^{\,2}
\left(Z\alpha/\beta p\right)^2f(\xi) .\end{equation}

The Coulomb correction to the potential (\ref{10}) reads
\begin{equation}
\Delta_{\scriptscriptstyle \mathrm{CC}} [U(\eta)] \equiv
U(\eta\,)-U^{\scriptscriptstyle \mathrm{B}}(\eta\,) =-4\pi
in^{}_0\eta^{\,2} \left(Z\alpha/\beta p\right)^2f(\xi)\
.\end{equation}

Now we obtain the corresponding relative Coulomb corrections. Using
(\ref{5}), we get
\begin{equation}
\delta_{\scriptscriptstyle \mathrm{CC}}\big[U(\eta)\big]\equiv
\frac{
 \Delta_{\scriptscriptstyle \mathrm{CC}} [U(\eta)]}{
U^{\scriptscriptstyle \mathrm{B}}(\eta\,)}
=\frac{\Delta_{\scriptscriptstyle \mathrm{CC}}
[\nu(\eta)]}{\nu^{\scriptscriptstyle \mathrm{B}}(\eta\,)}\equiv
\delta_{\scriptscriptstyle \mathrm{CC}}\big[\nu(\eta)\big]\ .
\end{equation}
Then (\ref{16}), (\ref{17}), and (\ref{nu}) give
\begin{eqnarray}\label{deltanu}
\delta_{\scriptscriptstyle \mathrm{CC}}\big[\nu(\eta)\big]=
\frac{f(Z\alpha/\beta)}{\ln\eta+\ln\left(\theta_a^{\scriptscriptstyle
\mathrm{B}}\right)-\ln 2+C_{\scriptscriptstyle \mathrm{E}}-0.5}
=-\frac{f(Z\alpha/\beta)}{0.615-\ln\left(1.2\,\alpha\,Z^{1/3}\right)-\ln\eta}\
.
\end{eqnarray}

We see from (\ref{deltanu}) and (\ref{22.7})  that
\begin{equation}
\delta_{\scriptscriptstyle
\mathrm{CC}}\big[\nu(\eta)\big]=\delta_{\scriptscriptstyle
\mathrm{CC}}\big[U(\eta)\big]<\delta_{\scriptscriptstyle
\mathrm{CC}}\big[\left\langle dI/d\omega\right\rangle\big]\ ,
\end{equation}
and we can estimate the $\delta_{\scriptscriptstyle
\mathrm{CC}}\big[\nu(\eta)\big]$ values using (\ref{deltanu})  for
$\eta\ll 1$. Their numerical values are presented in Table 5.

\begin{center} {\bf Table 5.}
The relative Coulomb corrections $\delta_{\scriptscriptstyle
\mathrm{CC}}\big[\nu(\eta)\big]$ and $\delta_{\scriptscriptstyle
\mathrm{CC}}\big[U(\eta)\big]$ for the gold, lead, and uranium
targets.
\end{center}

\begin{center}
\begin{tabular}{lccc}
\hline \\[-3mm]
Target&Z&~~~~$a\leq\eta\leq b$&~~~$ -\delta_{\scriptscriptstyle
\mathrm{CC}}\big[\nu(\eta)\big]=-\delta_{\scriptscriptstyle
\mathrm{CC}}\big[U(\eta)\big]$~~~\\[.2cm]
\hline \\[-3mm]
Au&79&~~~~~$0.01\leq \eta \leq 0.1$&~~~~~$3.7\%\leq
-\delta_{\scriptscriptstyle \mathrm{CC}}\big[\nu(\eta)\big]\leq
5.0\%$~~~\\
Pb&82&~~~~~$0.01\leq \eta \leq 0.1$&~~~~~$3.9\%\leq
-\delta_{\scriptscriptstyle
\mathrm{CC}}\big[\nu(\eta)\big]\leq 5.3\%$~~~\\
U&92&~~~~~$0.01\leq \eta \leq 0.1$&~~~~~$5.5\%\leq
-\delta_{\scriptscriptstyle
\mathrm{CC}}\big[\nu(\eta)\big]\leq 8.0\%$~~~\\[.2cm]

\hline\\[-3mm]
\end{tabular}
\end{center}
Thus, e.g., $-\delta_{\scriptscriptstyle
\mathrm{CC}}\big[\nu(\eta)\big]=-\delta_{\scriptscriptstyle
\mathrm{CC}}\big[U(\eta)\big]\sim 4.3\%<-\delta_{\scriptscriptstyle
\mathrm{CC}}\big[\left\langle dI/d\omega\right\rangle\big]\sim
8.0\%$ for $Z=79$ (Au).

\newpage

Let us consider the spectral bremsstrahlung intensity (\ref{6}) in
the form proposed by Migdal:
\begin{equation}\label{migdal}
\left\langle\frac{dI}{d\omega}\right\rangle
=\Phi(s)\left(\frac{dI}{d\omega}\right)_0\ ,
\end{equation}
where $(dI/d\omega)_0$ is the spectral bremsstrahlung rate without
accounting for the multiple scattering effects in the radiation,
\begin{equation}\label{w0}
\left(\frac{dI}{d\omega}\right)_0=\frac{2e^2}{3\pi} \gamma^2q\,L\ ,
\end{equation}
\begin{equation}\label{q}
q=\overline{\vartheta^2}/L\ .
\end{equation}
The function $\Phi(s)$ accounts for the multiple scattering
influence on the bremsstrahlung rate,
\begin{equation}\label{Phi}
\Phi(s)=24s^2\left[\int\limits_{0}^{\infty}
dx\,e^{-2sx}\mbox{cth}(x)\sin(2sx)-\frac{\pi}{4} \right],
\end{equation}
\begin{equation}\label{s2}
s^2=\lambda^2/\overline{\vartheta^2}\ .
\end{equation}
It has simple asymptotes at the small and large values of the
argument:
\begin{equation}\label{as}
\Phi(s) \rightarrow\left\{\begin{array}{cl}
6s,&s\;\rightarrow \;0\;,\\
1,&s\;\rightarrow \infty,\end{array}\right.
\end{equation}
\begin{equation}\label{s}
s=\frac{1}{4\gamma^2}\sqrt{\frac{\omega}{q}}\ .
\end{equation}
For $s\ll 1$, the suppression is large, and $\Phi(s)\approx 6s$. The
intensity of radiation in this case is much less, than the
corresponding result of Bethe and Heitler. If $s\geq 1$ (i.e.
$\omega \geq \omega_{cr}$), the function $\Phi(s)$ is close to a
unit,  and the following approximation is valid \cite{Sh1}:
\begin{equation}\label{small}
\Phi(s)\approx 1-0.012/s^4\ .
\end{equation}

The formula (\ref{migdal}) is obtained with the logarithmic
accuracy. At $s\gg 1$, (\ref{w0}) coincides to the logarithmic
accuracy with the Bethe--Heitler result
\begin{equation}\label{BH}
\left\langle\frac{dI}{d\omega}\right\rangle_{\scriptscriptstyle
 \mathrm{BH}} =\frac{L}{L_{\scriptscriptstyle
\mathrm{R}}}\left[1+\frac{1}{12\ln\left(183Z^{-1/3}\right)}\right]\
.
\end{equation}
If $s\ll 1$, we have the LPM suppression in comparison with
(\ref{BH}).

Now we obtain analytical and numerical results for the Coulomb
corrections to these quantities. In order to derive an analytical
expression for the Coulomb correction to the Born spectral
bremsstrahlung rate $(dI/d\omega)_0$, we first write
\begin{equation}\label{CCw0}
\Delta_{\scriptscriptstyle
\mathrm{CC}}\left[\left(\frac{dI}{d\omega}\right)_0\right]\equiv
\left(\frac{dI}{d\omega}\right)_0-
\left(\frac{dI}{d\omega}\right)^{\scriptscriptstyle \mathrm{B}}_0
=\frac{2e^2}{3\pi} \gamma^2L\cdot\Delta_{\scriptscriptstyle
\mathrm{CC}}[q]\ ,
\end{equation}
\begin{equation}\label{CCq}
\Delta_{\scriptscriptstyle \mathrm{CC}}[q]\equiv
q-q^{\scriptscriptstyle
\mathrm{B}}=\frac{1}{L}\cdot\Delta_{\scriptscriptstyle
\mathrm{CC}}\left[\overline{\vartheta^2}\right]\ .
\end{equation}
Accounting for $\overline{\vartheta^2}=\theta_c^2B$
(\ref{vartheta2}), we get
\begin{equation}\label{CCvartheta2}
\Delta_{\scriptscriptstyle
\mathrm{CC}}\left[\overline{\vartheta^2}\right]\equiv
\overline{\vartheta^2}-\left(\overline{\vartheta^2}\right)^{\scriptscriptstyle
\mathrm{B}}=\theta_c^2 \cdot\Delta_{\scriptscriptstyle
\mathrm{CC}}\left[B\right]\ .
\end{equation}
Then, using (\ref{b}) and (\ref{B}), we arrive at
\begin{equation}\label{limcorrect1}
\Delta_{\scriptscriptstyle \mathrm{CC}}[b]=-f(\xi)
 =\left(1-\frac{1}{B^{\scriptscriptstyle
\mathrm{B}}}\right)\cdot\Delta_{\scriptscriptstyle \mathrm{CC}}[B]\
 ,
\end{equation}
\begin{equation}\label{limcor1}
\Delta_{\scriptscriptstyle
\mathrm{CC}}[B]=\frac{f(\xi)}{1/B^{\scriptscriptstyle
\mathrm{B}}-1}\ .
\end{equation} In doing so, (\ref{CCw0}) becomes
\begin{equation}
\Delta_{\scriptscriptstyle
\mathrm{CC}}\left[\left(\frac{dI}{d\omega}\right)_0\right]
=\frac{2(e\gamma\theta_c)^2}{3\pi\,(1/B^{\scriptscriptstyle
\mathrm{B}}-1)}\cdot f(\xi)\ ,
\end{equation}
and the relative Coulomb correction reads

\begin{equation*}
\delta_{\scriptscriptstyle
\mathrm{CC}}\left[(dI/d\omega)_0\right]=\delta_{\scriptscriptstyle
\mathrm{CC}}\left[q\right]=\delta_{\scriptscriptstyle
\mathrm{CC}}\left[\overline{\vartheta^2}\right]=\delta_{\scriptscriptstyle
\mathrm{CC}}\left[B\right]
\end{equation*}
\begin{equation}\label{migdalparam}
=R_{\scriptscriptstyle
\mathrm{CC}}\left[(dI/d\omega)_0\right]-1=\frac{f(\xi)}{1-B^{\scriptscriptstyle
\mathrm{B}}}\ .
\end{equation}

Next, in order to obtain the relative Coulomb correction to the
Migdal function $\Phi(s)$, we first derive corresponding correction
to the quantity $s^2$ (\ref{s2}):
\begin{eqnarray}
\Delta_{\scriptscriptstyle \mathrm{CC}}\left[s^2
\right]&=&\frac{\omega}{16\gamma^4}
\left(\frac{1}{q}-\frac{1}{q^{\scriptscriptstyle
\mathrm{B}}}\right)\ ,\\
\label{CCs2} \delta_{\scriptscriptstyle \mathrm{CC}}\left[s^2
\right]&=&\frac{q^{\scriptscriptstyle
\mathrm{B}}}{q}-1=\frac{\big(\overline{\vartheta^2}\big)^{\scriptscriptstyle
\mathrm{B}}}{\overline{\vartheta^2}}-1
\end{eqnarray}
\begin{equation} =\frac{1}{\delta_{\scriptscriptstyle
\mathrm{CC}}\big[\overline{\vartheta^2}\big]+1}-1
=\frac{1}{R_{\scriptscriptstyle
\mathrm{CC}}\left[(dI/d\omega)_0\right]}-1\ .
\end{equation}
This leads to the following relative Coulomb correction for $s$
(\ref{s}):
\begin{equation}
\delta_{\scriptscriptstyle \mathrm{CC}}\left[s \right]
=\frac{1}{\sqrt{\delta_{\scriptscriptstyle
\mathrm{CC}}\big[\overline{\vartheta^2}\big]+1}}-1
=\frac{1}{\sqrt{R_{\scriptscriptstyle
\mathrm{CC}}\Big[(dI/d\omega)_0\Big]}}-1\ .
\end{equation}

For the asymptote $\Phi(s)=6s$ (\ref{as}), we get
\begin{equation}\label{CCPhi}
\delta_{\scriptscriptstyle \mathrm{CC}}\left[\Phi(s) \right]
=\delta_{\scriptscriptstyle \mathrm{CC}}\left[s \right]\ .
\end{equation}

Then, the total relative Coulomb correction to $\langle
dI/d\omega\rangle$ in this asymptotic case becomes:
\begin{equation}\label{sum}
\delta_{\scriptscriptstyle \mathrm{CC}}\left[\langle
dI/d\omega\rangle \right]=\delta_{\scriptscriptstyle
\mathrm{CC}}\left[(dI/d\omega)_0\right]+\delta_{\scriptscriptstyle
\mathrm{CC}}\left[\Phi(s) \right]\ .
\end{equation}

Numerical values of these corrections for some specified values of
the Moli\`{e}re parameter $B^{\scriptscriptstyle \mathrm{B}}$ are
presented in Table 6.

\smallskip

\begin{center} {\bf Table 6.}
Relative Coulomb corrections  to the parameters of the Migdal LPM
theory, $\delta_{\scriptscriptstyle
\mathrm{CC}}\left[(dI/d\omega)_0\right]$ (\ref{migdalparam}),
$\delta_{\scriptscriptstyle \mathrm{CC}}\left[\Phi(s) \right]$
(\ref{CCPhi}), and $\delta_{\scriptscriptstyle
\mathrm{CC}}\left[\langle dI/d\omega\rangle\right]$ (\ref{sum}), in
the regime of strong LPM suppression  for $Z=79$ (Au) and $\beta=1$.
\end{center}

\begin{center}
\begin{tabular}{ccccccc}
\hline \\[-3mm]
$B^{\scriptscriptstyle \mathrm{B}}$&$\delta_{\scriptscriptstyle
\mathrm{CC}}\left[\left(\frac{dI}{d\omega}\right)_{\scriptscriptstyle
0}\right]$ &$R_{\scriptscriptstyle
\mathrm{CC}}\left[\left(\frac{dI}{d\omega}\right)_{\scriptscriptstyle
0}\right]$ &$\delta_{\scriptscriptstyle \mathrm{CC}}\left[\Phi(s)
\right] $&$\delta_{\scriptscriptstyle
\mathrm{CC}}\left[\left\langle\frac{dI}{d\omega}\right\rangle\right]$&$R_{\scriptscriptstyle
\mathrm{CC}}\left[\left\langle\frac{dI}{d\omega}\right\rangle\right]$\\[.2cm]
\hline\\[-3mm]
4.50&$-0.089$&0.911&$-0.048$&$-0.137$&0.863\\
4.90&$-0.080$&0.920&$-0.043$&$-0.123$&0.877\\
8.46&$-0.042$&0.958&$-0.022$&$-0.064$&0.936\\[.2cm]
\hline
\end{tabular}
\end{center}

\smallskip

As can be seen from Table 6,  the moduli of the Coulomb corrections
to the quantities $(dI/d\omega)^{\scriptscriptstyle \mathrm{B}}_0$
and $\Phi^{\scriptscriptstyle \mathrm{B}}(s)$ decrease from about 9
to 4\% and from 5 to 2\%, respectively, with an increase in the
parameter $B^{\scriptscriptstyle \mathrm{B}}$ from a minimum value
$4.5$ \cite{C-7} to a value 8.46 corresponding to the conditions of
experiment \cite{Fomin}; and the modulus of the total relative
correction $\delta_{\scriptscriptstyle \mathrm{CC}}\left[\langle
dI/d\omega\rangle\right]$ decreases from approximately 14 to 6\%.

The average $\bar R_{\scriptscriptstyle \mathrm{CC}}=0.947\pm 0.015$
for the gold target at $B^{\scriptscriptstyle \mathrm{B}}=8.46$ from
Table 6 is close to the corresponding $\bar R_{\scriptscriptstyle
\mathrm{CC}}=0.945\pm 0.008$ from Table 4. This corresponds to the
mean value $\bar \delta_{\scriptscriptstyle \mathrm{CC}}=-5.4\%$,
which coincides with the value of the normalization correction
$-5.5\pm 0.2\%$ for $6\%L_{\scriptscriptstyle R}$ gold target (Table
II in \cite{C-1.1.2}).

A comparison of the non-averaged ratio value $R_{\scriptscriptstyle
\mathrm{CC}}\left[\left\langle
 dI/d\omega\right\rangle\right]=0.936$ from Table 6 with the
normalization factor $R\sim 0.94$ would be incorrect, because the
regime of strong suppression is not achieved in the analyzed SLAC
experiment. For such a comparison, we will carry out now calculation
for the regime of small LPM suppression \eqref{small}.

In order to obtain the relative correction
$\delta_{\scriptscriptstyle \mathrm{CC}}\left[\Phi(s) \right] $ in
this regime, we first derive an expression for the Coulomb
correction $\Delta_{\scriptscriptstyle \mathrm{CC}}\left[\Phi(s)
\right]$ to the Migdal function $\Phi(s)$:
\begin{equation}
\Delta_{\scriptscriptstyle \mathrm{CC}}\left[\Phi(s) \right]=0.012
\left(\frac{1}{\left(s^4\right)^{\scriptscriptstyle \mathrm{B}}}-
\frac{1}{s^4}\right)=\frac{0.012}{s^4}\,\delta_{\scriptscriptstyle
\mathrm{CC}}\left[s^4 \right]\ ,
\end{equation}
\begin{equation*}
\delta_{\scriptscriptstyle \mathrm{CC}}\left[s^4
\right]=\left(\frac{q^{\scriptscriptstyle
\mathrm{B}}}{q}\right)^2-1=\left(\frac{\big(\overline{\vartheta^2}\big)^{\scriptscriptstyle
\mathrm{B}}}{\overline{\vartheta^2}}\right)^2-1=1/\left(\delta_{\scriptscriptstyle
\mathrm{CC}}\big[\overline{\vartheta^2}\big]+1\right)^2-1
\end{equation*}
\begin{equation}\label{CCs4}
=1/\Big(R_{\scriptscriptstyle
\mathrm{CC}}\left[(dI/d\omega)_0\right]\Big)^2-1\ .
\end{equation}

This leads to the following relative Coulomb correction for
$\Phi(s)$ (\ref{small}):
\begin{equation}\label{CCPhismall}
\delta_{\scriptscriptstyle \mathrm{CC}}\left[\Phi(s) \right]
=\frac{0.012}{s^4}\,\delta_{\scriptscriptstyle \mathrm{CC}}\left[s^4
\right] \cdot\frac{\left(s^4\right)^{\scriptscriptstyle \mathrm{B}}
}{\left(s^4\right)^{\scriptscriptstyle \mathrm{B}}-0.012}
=0.012\,\frac{\delta_{\scriptscriptstyle \mathrm{CC}}\left[s^4
\right]}{\delta_{\scriptscriptstyle \mathrm{CC}}\left[s^4 \right]+1}
\cdot\frac{1}{\left(s^4\right)^{\scriptscriptstyle
\mathrm{B}}-0.012}\ .
\end{equation}

In Table 7 are listed the values of the relative Coulomb corrections
to the quantities of \eqref{migdal} in the regime of small
suppression (\ref{small}) for some separate $s$ values (s=1.2,
s=1.3).

\smallskip

\begin{center}
{\bf Table 7.} Relative Coulomb corrections  to the quantities of
the Migdal LPM theory, $\delta_{\scriptscriptstyle
CC}\left[(dI/d\omega)_0\right]$ (\ref{migdalparam}),
$\delta_{\scriptscriptstyle CC}\left[\Phi(s) \right]$
(\ref{CCPhismall}), and $\delta_{\scriptscriptstyle CC}\left[\langle
dI/d\omega\rangle\right]$ (\ref{sum}), in the regime of small LPM
suppression for high Z targets of experiment \cite{C-1.1.2}
\medskip

{\bf 1.} for $\beta=1$, $B^{\scriptscriptstyle B}=8.46$, $s=1.2$

\smallskip

\begin{tabular}{ccccccc}
\hline \\[-3mm]
Target &Z&$\delta_{\scriptscriptstyle
CC}\left[\left(\frac{dI}{d\omega}\right)_{\scriptscriptstyle
0}\right]$ &$\delta_{\scriptscriptstyle \mathrm{CC}}\left[s^4
\right]$&$\delta_{\scriptscriptstyle CC}\left[\Phi(s) \right]
$&$\delta_{\scriptscriptstyle
CC}\left[\left\langle\frac{dI}{d\omega}\right\rangle\right]$&$R_{\scriptscriptstyle
CC}\left[\left\langle\frac{dI}{d\omega}\right\rangle\right]$\\[.2cm]
\hline\\[-3mm]
Au&79&$-0.0420$&$-0.0896$&$-0.0006$&$-0.0426$&0.9574\\
Pb&82&$-0.0445$&$-0.0953$&$-0.0006$&$-0.0451$&0.9549\\
U~ &92&$-0.0529$&$-0.1149$&$-0.0007$&$-0.0536$&0.9464\\[.2cm]
\hline
\end{tabular}

\smallskip

$\bar{R}_{\scriptscriptstyle CC}\left[\left\langle
dI/d\omega\right\rangle\right]=0.953\pm 0.006;\qquad
\bar{\delta}_{\scriptscriptstyle CC}\left[\left\langle
dI/d\omega\right\rangle\right]= -4.71\pm 0.58\%$.

\medskip

{\bf 2.} for $\beta=1$, $B^{\scriptscriptstyle B}=8.46$, $s=1.3$

\smallskip

\begin{tabular}{ccccccc}
\hline \\[-3mm]
Target &Z&$\delta_{\scriptscriptstyle
CC}\left[\left(\frac{dI}{d\omega}\right)_{\scriptscriptstyle
0}\right]$ &$\delta_{\scriptscriptstyle \mathrm{CC}}\left[s^4
\right]$&$\delta_{\scriptscriptstyle CC}\left[\Phi(s) \right]
$&$\delta_{\scriptscriptstyle
CC}\left[\left\langle\frac{dI}{d\omega}\right\rangle\right]$&$R_{\scriptscriptstyle
CC}\left[\left\langle\frac{dI}{d\omega}\right\rangle\right]$\\[.2cm]
\hline\\[-3mm]
Au&79&$-0.0420$&$-0.0896$&$-0.0004$&$-0.0424$&0.9576\\
Pb&82&$-0.0445$&$-0.0953$&$-0.0004$&$-0.0449$&0.9551\\
U~ &92&$-0.0529$&$-0.1149$&$-0.0005$&$-0.0534$&0.9466\\[.2cm]
\hline
\end{tabular}

\smallskip

$\bar{R}_{\scriptscriptstyle CC}\left[\left\langle
dI/d\omega\right\rangle\right]=0.953\pm 0.006;\qquad
\bar{\delta}_{\scriptscriptstyle CC}\left[\left\langle
dI/d\omega\right\rangle\right]= -4.69\pm 0.58\%$.

\end{center}

Figure 1 demonstrates the $s$ dependence of the corrections
$-\bar{\delta}_{\scriptscriptstyle CC}\left[\left\langle
dI/d\omega\right\rangle\right]\,(\%)$ over the entire range $1.0
\leq s\leq \infty$ of the parameter $s$, for which the regime of
small LPM suppression is valid. Their sampling mean over this range
$\bar{\delta}_{\scriptscriptstyle CC}\left[\left\langle
dI/d\omega\right\rangle\right]= -4.70\pm 0.49\%$ gives Table 8. The
asymptotic value of $\bar{\delta}_{\scriptscriptstyle
CC}\left[\left( dI/d\omega\right)_0\right]$ is $-4.65\pm 0.45\%$.

\newpage

\begin{figure}[h!]

\begin{center}

\includegraphics[width=0.65\linewidth]{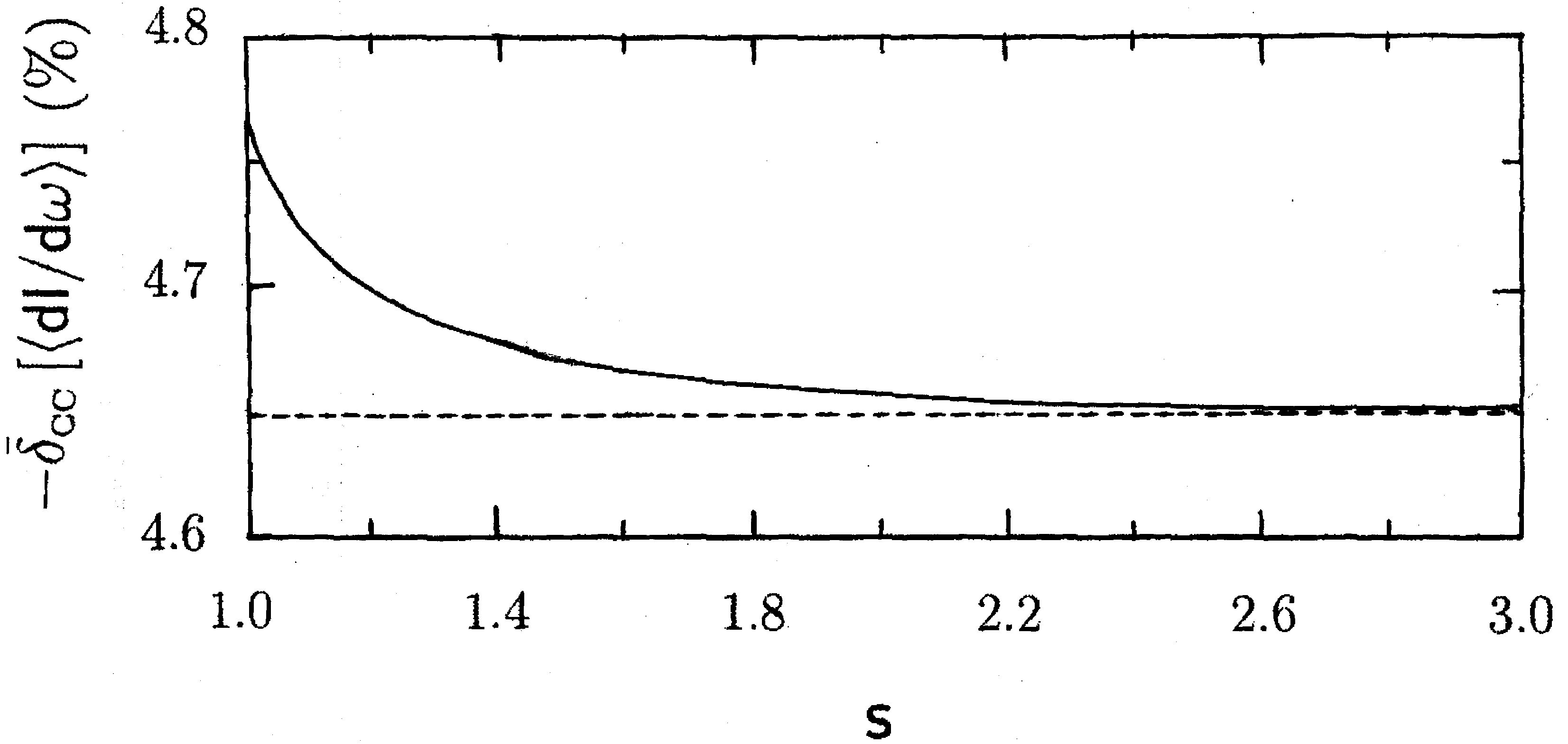}
\caption{The $s$ dependence of the corrections
$-\bar{\delta}_{\scriptscriptstyle CC}\left[\left\langle
dI/d\omega\right\rangle\right]\,(\%)$ over the entire range $1.0 \leq
s\leq \infty$ of the parameter $s$.\label{}}

\end{center}

\end{figure}

Table 8 presents the values of the corrections
$-{\delta}_{\scriptscriptstyle CC}\left[\left\langle
dI/d\omega\right\rangle\right]\,(\%)$ for some separate high Z
target elements over the range  $1.0 \leq s\leq \infty$.

\medskip

\begin{center}
{\bf Table 8.} The dependence of $-\delta_{\scriptscriptstyle
CC}\left[\langle dI/d\omega\rangle\right]$ values on the parameter
$s$ in the regime of small LPM suppression for some high Z targets
of experiment \cite{C-1.1.2} at $\beta=1$ and $B^{\scriptscriptstyle
B}=8.46$.

\medskip

\begin{tabular}{ccccccccc}
\hline \\[-2mm]
Target&$Z$&s=1.0 &s=1.1 &s=1.2 &s=1.3&s=1.5&s=2.0&s=$\infty$\\[.2cm]
\hline\\[-3mm]
Au&79&0.0432&0.0428&0.0426&0.0424&0.0422&0.0421&0.0420\\
Pb&82&0.0458&0.0454&0.0451&0.0449&0.0447&0.0446&0.0445\\
U~~&92&0.0545&0.0540&0.0536&0.0534&0.0532&0.0530&0.0529\\[.2cm]
\hline
\end{tabular}

\medskip

$\delta_{\scriptscriptstyle
CC}\left[\left\langle\frac{dI}{d\omega}\right\rangle\right]$=$-4.50\pm
0.05\%$ ($Z$=$82$), \, $\delta_{\scriptscriptstyle
CC}\left[\left\langle\frac{dI}{d\omega}\right\rangle\right]$=$-5.35\pm
0.06\%$ ($Z$=$92$),\\
\smallskip
 $\bar{\delta}_{\scriptscriptstyle
CC}\left[\left\langle
dI/d\omega\right\rangle\right]= -4.70\pm 0.49\%$.\\

\end{center}

\medskip

Table 8  shows that averaging over the range $1.0 \leq s\leq \infty$
corrections $\delta_{\scriptscriptstyle CC}\left[\langle
dI/d\omega\rangle\right]$  for some separate high Z
targets\footnote{For low Z targets, the E-146 data showed a
disagreement with the Migdal LPM theory predictions. There is a
problem of an adequate describe the photon spectra shape for the low
Z targets \cite{C-1.1.2,Klein3}. Therefore, we will analyze only
results for some high Z targets of the SLAC E-146 experiment.} gives
their sampling means $\delta_{\scriptscriptstyle CC}\left[\langle
dI/d\omega\rangle\right]=-4.50\pm 0.05\%$ ($Z=82$) and
$\delta_{\scriptscriptstyle CC}\left[\langle
dI/d\omega\rangle\right]=-5.35\pm 0.06\%$ ($Z=92$), which coincide
with the normalization correction values $-4.5\pm 0.2\%$ for
$2\%L_{\scriptscriptstyle R}$ lead target and $-5.6\pm 0.3\%$ for
$3\%L_{\scriptscriptstyle R}$ uranium target (Table II in
\cite{C-1.1.2}), respectively, within the experimental error.

Averaging over this range corrections
$\bar{\delta}_{\scriptscriptstyle CC}\left[\left\langle
dI/d\omega\right\rangle\right]$ gives the sampling mean $-4.70\pm
0.49\%$,  which excellent agrees with the  weighted average value
$-4.7\pm 2\%$ of the normalization correction obtained in
\cite{C-1.1.2} for 25 GeV data\footnote{It becomes $-4.8\pm 3.5\%$
for the 8 GeV data if the outlying $6\%L_{\scriptscriptstyle R}$
gold target is excluded from them \cite{C-1.1.2}.}. We believe that
this allows one to understand the origin of the discussed in
\cite{C-1.1.1,C-1.1.2} normalization problem for high Z targets.

\subsection{\normalsize{Fokker--Planck
approximation accuracy in the case $\boldsymbol\omega\mathbf{=0}$ }}

Finally, let us briefly discuss the accuracy of the Fokker--Planck
approximation that allows to obtain an analytical expression to be
derived for the Migdal particle distribution function and the entire
$\left\langle dI(\omega)/d\omega\right\rangle$ range to be  rather
simply calcula\-te (using numerical calculation of triple
integrals).

To this end, we will fix the parameter  $a$ in expression (\ref{10})
in such a way that the results of the exact calculation of
$\left\langle dI(\omega)/d\omega\right\rangle\big\vert_{\omega
\gg\omega_{cr}}$
 and its calculation in the Fokker--Planck approximation coincide.
As a result, we get

\begin{equation}\label{26}
a= 2\pi\biggl (\frac{Z\alpha\sigma}{m} \biggl )^2 \biggl (\ln
\frac{\sigma}{\theta_a}+ \frac{7}{12}\biggl )\ .
\end{equation}

Now we calculate $\left\langle
dI(\omega)/d\omega\right\rangle\big\vert_{\omega = 0}$ using the
relations (\ref{10}) and (\ref{26}) and compare the result with the
result obtained using `realistic' (Moli\`{e}re) expression
(\ref{16}) for $\nu(\eta)$. Then for the ratio
\begin{equation}\label{27}
R_{\scriptscriptstyle \mathrm{FPM}} = \frac{\left\langle
dI(\omega)/d\omega
 \right\rangle_{\scriptscriptstyle \mathrm{FP}}}
{\left\langle dI(\omega)/d\omega
 \right\rangle_{\scriptscriptstyle \mathrm{M}}}
\end{equation}
we get the following values:
\begin{equation}\label{28} R_{\scriptscriptstyle \mathrm{FPM}}(\omega,L)=
\left\{\begin{array}{cc} 0.890,&L=0.007\,L_{\scriptscriptstyle
\mathrm{R}}\\
0.872,&L=0.060\,L_{\scriptscriptstyle \mathrm{R}}\end{array}\right.\
.
\end{equation}

The values of corresponding relative corrections
\begin{equation}
\delta_{\scriptscriptstyle \mathrm{FPM}}\big[\!\left\langle
dI/d\omega\right\rangle\!\big]=\frac{\left\langle
dI(\omega)/d\omega\right\rangle_{\scriptscriptstyle \mathrm{FP}}-
\left\langle dI(\omega)/d\omega \right\rangle_{\scriptscriptstyle
\mathrm{M}}}{\left\langle dI(\omega)/d\omega
\right\rangle_{\scriptscriptstyle \mathrm{M}}}
\end{equation}
in percentage are given in Table 9.

\begin{center} {\bf Table 9.}
The relative correction $\delta_{\scriptscriptstyle
\mathrm{FPM}}\big[\left\langle
 dI/d\omega\right\rangle \big]$ for $Z=79$ and $\omega=0$.
\end{center}

\begin{center}
\begin{tabular}{ccc}
\hline \\[-3mm]
\!L(cm)&$-\delta_{\scriptscriptstyle \mathrm{FPM}}\big[\left\langle
 dI/d\omega\right\rangle \big]
\!$&$\!R_{\scriptscriptstyle
 \mathrm{FPM}}\big[\left\langle
 dI/d\omega\right\rangle \big]\!$\\[.2cm]
\hline\\[-3mm]
$\!0.007\,L_{\scriptscriptstyle
\mathrm{R}}$&~~~~~0.110\!&\!0.890\!\\
$\!0.060\,L_{\scriptscriptstyle \mathrm{R}}$&~~~~~0.128\!&\!0.872\!\\[.2cm]
\hline
\end{tabular}
\end{center}

\smallskip

It is obvious that the relative difference between the
Fokker--Planck appro\-xi\-mation and the description based on the
Moli\`{e}re theory $\delta_{\scriptscriptstyle
\mathrm{FPM}}\big[\left\langle
 dI/d\omega\right\rangle \big]$ is about 12\%, which is noticeably higher than the
$3.2\%$ characteristic systematic experimental error \cite{C-1.1.1}.

Thus, the Fokker--Planck approximation and Gaussian distribution can
not be used  for describing the experimental data
\cite{C-1.1.1,C-1.1.2} at low frequencies $\omega < 30$ MeV. For
their description the application of the Moli\`{e}re multiple
scattering theory is advisable.

\subsection{\normalsize\textbf{Coulomb corrections in the
LPM effect theory analogue for a thin target}}

In \cite{Fomin} it is shown that the region of the emitted photon
frequencies $\omega_{cr}> \omega > 0$ naturally splits into two
intervals, $\omega_{cr} > \omega > \omega_{c}$ and $\omega_{c} >
\omega > 0$, in first of which the LPM effect for sufficiently thick
targets takes place, and in the second, there is its analogue for
thin targets. The quantity $\omega_{c}$ is defined here as
$\omega_{c}=2E^2/(m^2L)$.

Application of the Moli\`{e}re multiple scattering theory to the
analysis of experimental data \cite{C-1.1.1,C-1.1.2} for a thin
target in the second $\omega$ range is based on the use of the
expression for the spatial-angle particle distribution function
(\ref{14}) which satisfies the standard Boltzmann transport equation
for a thin homogeneous foil, and it differs significantly from the
Gaussian particle distri\-bu\-tion of the Migdal LPM effect theory.

Besides, it determines another expression for the spectral radiation
rate in the context of the coherent radiation theory
\cite{Fomin}\footnote{Note that the authors of \cite{Fomin} neglect
the influence of the medium polarization \cite{TM} on the radiation
in this theory. This is admissible in the conditions of the
experiment \cite{C-1.1.1,C-1.1.2}, where the LPM effect is more
important for photon energies above 5 MeV (at 25 GeV beams); and
dielectric suppression dominates at significantly lower photon
energies.}, which reads
\begin{equation}\label{averaged}
\,\left\langle\frac{dI}{d\omega}\right\rangle = \int
w_{\scriptscriptstyle
\mathrm{M}}(\vartheta)\frac{dI(\vartheta)}{d\omega}d^2\vartheta\ .
\end{equation}\\
Here
\begin{equation}\label{nonaveraged}
\frac{dI(\vartheta)}{d\omega}=\frac{2e^2}{\pi}
\left[\frac{2\chi^2+1}{\chi\sqrt{\chi^2+1}}
\ln\left(\chi+\sqrt{\chi^2+1}\right)-1\right]
\end{equation}
with $\chi=\gamma\vartheta/2$. The latter expression is valid for
consideration of the particle scattering in both amorphous and
crystalline medium.

The formula (\ref{nonaveraged}) has simple asymptotes at the small
and large values of  parameter $\chi=\gamma\vartheta/2$:
\begin{equation}\label{asymptnon}
\frac{dI(\vartheta)}{d\omega}
=\frac{2e^2}{3\pi}\left\{\begin{array}{cl}
\gamma^2\vartheta^2,&\gamma\vartheta\ll 1\;,\\
3\left[\ln (\gamma^2\vartheta^2)-1\right],&\gamma\vartheta\gg
1\;,\end{array}\right.
\end{equation}

Replacing $\vartheta^2$  by the average square value of the
scattering angle $\overline{\vartheta^2}$ in this formula, we arrive
at the following estimates for the average radiation spectral
density:
\begin{equation}\label{asymptaver}
\left\langle\frac{dI}{d\omega}\right\rangle
=\frac{2e^2}{3\pi}\left\{\begin{array}{cl}
\gamma^2\overline{\vartheta^2},&\gamma^2\overline{\vartheta^2}\ll 1\;,\\
3\left[\ln
(\gamma^2\overline{\vartheta^2})-1\right],&\gamma^2\overline{\vartheta^2}\gg
1\;.\end{array}\right.
\end{equation}

In the experiment  \cite{C-1.1.1,C-1.1.2},  the above  frequency
intervals correspond roughly to the following $\omega$ ranges:
$(\omega_{cr}
> \omega
> \omega_{c})\sim (244\,\mbox{MeV} > \omega
> 30\,\mbox{MeV})$ and $(\omega_{c} > \omega > 0)\sim (30\,\mbox{MeV} > \omega
> 5\,\mbox{MeV})$ for 25 GeV electron beam and
$0.7-6.0\%L_{\scriptscriptstyle \mathrm{R}}$ gold target.  Whereas
in the first area the discrepancy between the LPM theory predictions
and data is about 3.2 to 5\%, in the second area this discrepancy
reaches $\sim 15\%$.

Using the approximate second-order representation of the Moli\`{e}re
distri\-bu\-tion function (\ref{2order}), (\ref{W_1}) for computing
the spectral radiation rate (\ref{averaged}) the authors of
\cite{Fomin} succeeded to agree satisfactorily theory and 25 GeV and
$0.7\%L_{\scriptscriptstyle \mathrm{R}}$ data over the $\omega$
range 5 to 30 MeV.

This result can be understood by considering the fact that the
correction to the Gaussian first-order representation of the
distribution function $w_{\scriptscriptstyle \mathrm{M}}(\vartheta)$
of order of $1/B^{\scriptscriptstyle \mathrm{B}}$ is about 12\% for
the value used in calculations $B^{\scriptscriptstyle
\mathrm{B}}=8.46$ \cite{Fomin}.

Let us obtain the relative Coulomb correction to the averaged value
of the spectral density of radiation for two limiting cases
(\ref{asymptaver}).

In the first case $\gamma^2\overline{\vartheta^2}\ll 1$, taking into
account the equality
\begin{equation}
\delta_{\scriptscriptstyle
\mathrm{CC}}[\gamma^2\overline{\vartheta^2}]=\delta_{\scriptscriptstyle
\mathrm{CC}}[\overline{\vartheta^2}]\ ,
\end{equation}
(\ref{migdalparam}),  and (\ref{asymptaver}), we get
\begin{equation}\label{limcorrect1}
\delta_{\scriptscriptstyle \mathrm{CC}}\left[\left\langle
\frac{dI}{d\omega}\right\rangle\right]=\delta_{\scriptscriptstyle
\mathrm{CC}}\left[\left( \frac{dI}{d\omega}\right)_0\right]
=\frac{f(\xi)}{1-B^{\scriptscriptstyle \mathrm{B}}}\ ,
\end{equation}
where $B^{\scriptscriptstyle \mathrm{B}} \approx 8.46$ in the
conditions of the discussed experiment \cite{Fomin}.

In the second case $\gamma^2\overline{\vartheta^2}\gg 1$, we have
\begin{equation}
\Delta_{\scriptscriptstyle \mathrm{CC}}\left[\ln\left(
\gamma^2\overline{\vartheta^2}\right)-1\right]=\Delta_{\scriptscriptstyle
\mathrm{CC}}\left[\ln\left(\overline{\vartheta^2}\right)\right]
=\Delta_{\scriptscriptstyle \mathrm{CC}}\big[\ln\left(B\right)\big]\
.
\end{equation}
For the latter quantity, one can obtain
\begin{equation}
\Delta_{\scriptscriptstyle
\mathrm{CC}}[\ln\left(B\right)]=\Delta_{\scriptscriptstyle
\mathrm{CC}}[B]+f(Z\alpha)= \delta_{\scriptscriptstyle
\mathrm{CC}}[B]\ .
\end{equation}
The Coulomb correction then becomes
\begin{equation}
\Delta_{\scriptscriptstyle \mathrm{CC}}\left[\ln\left(
\gamma^2\overline{\vartheta^2}\right)-1\right]=\frac{\delta_{\scriptscriptstyle
\mathrm{CC}}[B]}{\left[\ln
(\gamma^2\overline{\vartheta^2})^{\scriptscriptstyle
\mathrm{B}}-1\right]}\ .
\end{equation}
Taking into account (\ref{migdalparam}), we arrive at a result:
\begin{equation}\label{limcorrect2}
\delta_{\scriptscriptstyle \mathrm{CC}}\left[\left\langle
\frac{dI}{d\omega}\right\rangle\right]= \frac{f(\xi)}{\left[\ln
(\gamma^2\overline{\vartheta^2})^{\scriptscriptstyle
\mathrm{B}}-1\right]\Big(1-B^{\scriptscriptstyle \mathrm{B}}\Big)}\
.
\end{equation}
The numerical values of these corrections are presented in Table 10.

\begin{center} {\bf Table 10.}
The relative Coulomb correction $\delta_{\scriptscriptstyle
\mathrm{CC}}\big[\left\langle dI/d\omega\right\rangle \big]$ to the
asymptotes of the Born spectral radiation rate over the range
$\omega<\omega_{c}$ for $\beta=1$, $B^{\scriptscriptstyle
\mathrm{B}}\approx 8.46$, and
$\left(\gamma^2\overline{\vartheta^2}\right)^{\scriptscriptstyle
\mathrm{B}}\approx 7.61$ \cite{Fomin}.
\end{center}

\begin{center}
\begin{tabular}{lcccc}
\hline \\[-3mm]
Target&Z&$\gamma^2\overline{\vartheta^2}$&$-\delta_{\scriptscriptstyle
\mathrm{CC}}\big[\left\langle dI/d\omega\right\rangle\big]$&
$R_{\scriptscriptstyle \mathrm{CC}}
$\\[.2cm]
\hline\\[-3mm]
Au&79&$\gamma^2\overline{\vartheta^2}\ll 1$&$0.042$&0.958\\
Au&79&$\gamma^2\overline{\vartheta^2}\gg 1$&$0.040$&0.960\\[.2cm]
\hline
\end{tabular}
\end{center}

\smallskip

The second asymptote  is not reached \cite{Fomin} in the experiment
\cite{C-1.1.1, C-1.1.2}. Therefore, we will now consider another
limiting case corresponding to the experimental conditions  and
taking into account the second term of the Moli\`{e}re distribution
function expansion (\ref{power}).

Substituting the second-order expression (\ref{2order}) for the
distribution function  in (\ref{averaged}) and integrating its
second term (\ref{W_1}), we can arrive at the following expression
for the electron radiation  spectrum at
$\mu^2=\gamma^2\overline{\vartheta^2}\gg 1$ \cite{Fomin}:

\begin{equation}\label{comleteasymp}
\left\langle\frac{dI}{d\omega}\right\rangle =\frac{2e^2}{\pi}
\left\{\ln \left(\mu^2 \right)-C_{\scriptscriptstyle
\mathrm{E}}\left(1+\frac{2}{\mu^2}\right)+\frac{2}{\mu^2}+\frac{C_{\scriptscriptstyle
\mathrm{E}}}{B}-1\right\}\ .
\end{equation}

In order to obtain the Coulomb correction to the Born spectral
radiation rate from (\ref{comleteasymp}), we first calculate its
numerical value at $(\mu^2)^{\scriptscriptstyle \mathrm{B}}\approx
7.61$ and $B^{\scriptscriptstyle \mathrm{B}}\approx 8.46$, and we
become $\left\langle dI/d\omega\right\rangle^{\scriptscriptstyle
\mathrm{B}}=0.00542$. The Bethe--Heitler formula in the Born
approximation gets $\left\langle
dI/d\omega\right\rangle_{\scriptscriptstyle
\mathrm{BH}}^{\scriptscriptstyle \mathrm{B}}=0.00954$.

Then, we calculate the numerical values of $B$ and $\mu^2$
parameters including the Coulomb corrections. From
\begin{equation}
\Delta_{\scriptscriptstyle
\mathrm{CC}}[B]=\frac{f(\xi)}{1/B^{\scriptscriptstyle \mathrm{B}}-1}
=-0.355\ ,
\end{equation}\\
we obtain $B\approx 8.105$ for $Z=79$ and $B^{\scriptscriptstyle
\mathrm{B}}\approx 8.46$. The equality
\begin{equation}\label{mu}
\Delta_{\scriptscriptstyle \mathrm{CC}}\left[\ln \mu^2\right]
=\Delta_{\scriptscriptstyle \mathrm{CC}}\left[\ln
B\right]=\Delta_{\scriptscriptstyle
\mathrm{CC}}[B]+f(\xi)=\delta_{\scriptscriptstyle \mathrm{CC}}[B]
=-0.042
\end{equation}
gets $\ln \mu^2=1.987$ and $\mu^2=7.295$. Substituting these values
in (\ref{comleteasymp}), we have $\left\langle
dI/d\omega\right\rangle =0.00531$. The relative Coulomb corrections
to these parameters are presented in Table 11. These corrections are
not large. Their sizes are between two to four percent, i.e., of
order of the experimental error.

\begin{center} {\bf Table 11.}
The relative Coulomb corrections in the analogue of the LPM effect
theory for $0.07\,L_{\scriptscriptstyle \mathrm{R}}$ gold target,
$\omega<\omega_{c}$, and $\beta=1$.
\end{center}
\begin{center}
\begin{tabular}{ccccc}
\hline \\[-3mm]
$\delta_{\scriptscriptstyle \mathrm{CC}}[B]$&
$\delta_{\scriptscriptstyle \mathrm{CC}}\left[\ln\mu^2 \right] $
&$\delta_{\scriptscriptstyle
\mathrm{CC}}\left[\left(dI/d\omega\right)_{\scriptscriptstyle
0}\right]$ &$\delta_{\scriptscriptstyle
\mathrm{CC}}\big[\left\langle dI/d\omega\right\rangle\big]$
&$\delta_{\scriptscriptstyle \mathrm{CC}}\left[\Phi(s)\right]$\\[.2cm]
\hline\\[-3mm]
$-0.042$&$-0.021$&$-0.042$&$-0.020$&$-0.021$\\[.2cm]
\hline
\end{tabular}
\end{center}

\smallskip

Accounting for the relative Coulomb correction to the Bethe--Heitler
spectrum of brems\-strahlung, we find
$\left(dI/d\omega\right)_{\scriptscriptstyle \mathrm{BH}} =0.00916$.
So we get

\begin{equation}\label{migdnum}
\left\langle\frac{dI}{d\omega}\right\rangle
=0.580\left(\frac{dI}{d\omega}\right)_{\scriptscriptstyle
\mathrm{BH}}\ .
\end{equation}

This leads to the value of the spectral radiation rate in terms of
$dN/[d(\log \omega)]$ $\times 1/L_{\scriptscriptstyle \mathrm{R}}$,
where $N$ is the number of events per photon energy bin per incident
electron,
 $dN/[d(\log \omega)/L_{\scriptscriptstyle
\mathrm{R}}]=0.118\times 0.580=0.068$, which agrees very well with
the experimental result over the frequency range $\omega<30$ MeV for
25 GeV and $0.7\%L_{\scriptscriptstyle \mathrm{R}}$ gold target.
This result additionally improves the agreement between the theory
and experiment (see Fig. 2). It is close to the Zakharov result
\cite{Z2} and coincides with the result of Blancenbeckler and Drell
obtained in the eikonal approximation, which excellent agrees with
$0.7\%L_{\scriptscriptstyle \mathrm{R}}$ 25 GeV data for $\omega >
5$ MeV (see Figs. 12a in \cite{C-1.1.2} and 20a in \cite{Klein3}).

\begin{figure}[h!]

\begin{center}

\includegraphics[width=0.5\linewidth]{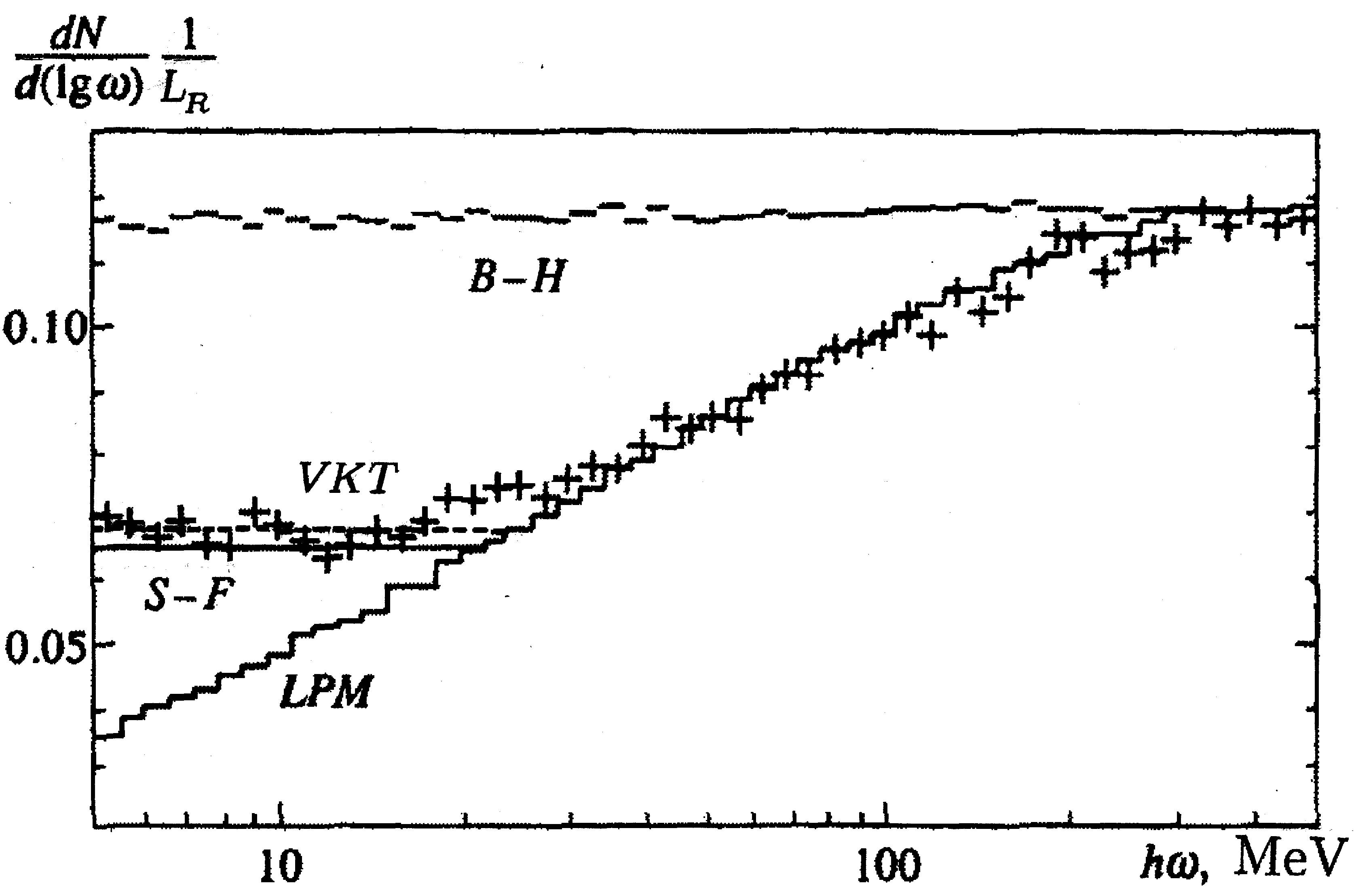}

\caption{\footnotesize Measurement of the LPM effect over the range
$30<\omega< 500$ MeV and its analogue in the range $5<\omega < 30$
MeV for the $0.7\%L_{\scriptscriptstyle R}$ gold target and 25 GeV
electron beam. The signs `+' denote the experimental data; the
histograms B--H and LPM give the Bethe--Heitler and the LPM Monte
Carlo predictions \cite{C-1.1.1}. The solid and dashed lines (S--F
and VKT) over the range $\omega < 30$ MeV are the results of
calculations without \cite{Fomin} and with the obtained Coulomb
corrections. } \label{Fig1}

\end{center}

\end{figure}

\section*{\Large{Summary and conclusions}}

\begin{itemize}

\item
Within the theory of LPM effect for finite-size targets, we
calculated the Coulomb corrections to the Born bremsstrahlung rate
and estimated the ratio $\left\langle
dI(\omega)/d\omega\right\rangle/\left\langle dI(\omega)/d\omega
\right\rangle^{\scriptscriptstyle \mathrm{B}}=R(\omega,L)$ for the
gold target based on results of the revised Moli\`{e}re  multiple
scattering theory for the Coulomb corrections to the screening
angle.

\item
We demonstrated that this $R(\omega,L)$ value
is close to the normalization constant $R$ value for
$0.7-6\%L_{\scriptscriptstyle \mathrm{R}}$ (25 GeV) data over the
$\omega$ range 30 to 500 MeV from \cite{C-3,C-1.1.1}; however, the
latter ignores the dependence of the ratio on $\omega$ and $L$.

\item
We obtained the analytical and numerical results for the Coulomb
corrections to the function
$\nu(\eta)=2\pi\int\sigma_0(\boldsymbol{\theta})[1-J_0(\eta\theta)]\boldsymbol{\theta}
d\boldsymbol{\theta}$  and complex potential $U(\eta)=-
\omega\lambda^2/2 - i\,n^{}_0 \nu(\eta)$, and we showed that
$-\delta_{\scriptscriptstyle
\mathrm{CC}}\big[\nu(\eta)\big]=-\delta_{\scriptscriptstyle
\mathrm{CC}}\big[U(\eta)\big]\sim 4.3\%<-\delta_{\scriptscriptstyle
\mathrm{CC}}\big[\left\langle dI/d\omega\right\rangle\big]\sim
8.0\%$ for $Z=79$ ($\beta=1$).

\item
Additionally, we found Coulomb corrections to the quantities of the
classical Migdal LPM theory, i.e., $\Delta_{\scriptscriptstyle
\mathrm{CC}}\!\left[s\right]$, $\Delta_{\scriptscriptstyle
\mathrm{CC}}\!\left[s^2\right]$, $\Delta_{\scriptscriptstyle
\mathrm{CC}}\!\left[s^4\right]$, $\Delta_{\scriptscriptstyle
\mathrm{CC}}\left[q\right]$, $\Delta_{\scriptscriptstyle
\mathrm{CC}}\left[(dI/d\omega)_0\right]$,
$\Delta_{\scriptscriptstyle \mathrm{CC}}\left[\Phi(s)\right]$,
$\Delta_{\scriptscriptstyle \mathrm{CC}}\left[\langle
dI/d\omega\rangle \right]$.

\item We calculated relative Coulomb corrections
$\delta_{\scriptscriptstyle
\mathrm{CC}}\left[(dI/d\omega)_0\right]=\delta_{\scriptscriptstyle
\mathrm{CC}}\left[q\right]$, $\delta_{\scriptscriptstyle
\mathrm{CC}}\left[\Phi(s) \right] =\delta_{\scriptscriptstyle
\mathrm{CC}}\left[s \right]$, and $\delta_{\scriptscriptstyle
\mathrm{CC}}\left[\langle dI/d\omega\rangle \right]$ in the regime
of strong LPM suppression for $Z=79$ ($\beta=1$).  We showed that
the latter correction $\delta_{\scriptscriptstyle
\mathrm{CC}}\left[\langle dI/d\omega\rangle \right]$ comprises the
order of $-14\%$ at minimum $B^{\scriptscriptstyle \mathrm{B}}$
value 4.5.

\item
We demonstrated that the average value $-5.4\%$ of the relative
Coulomb correction for $Z=79$ coincides with the normalization
correction value $-5.5\pm 0.2\%$ for $6\%L_{\scriptscriptstyle R}$
gold target obtained  in experiment \cite{C-1.1.2}.

\item
We have performed the analogous calculations for the regime of small
LPM suppression over the entire range $1\leq s \leq \infty$, and  we
found that the  values of the Coulomb corrections
$\delta_{\scriptscriptstyle CC}\left[\langle
dI/d\omega\rangle\right]=-4.50\pm 0.05\%$ ($Z=82$) and
$\delta_{\scriptscriptstyle CC}\left[\langle
dI/d\omega\rangle\right]=-5.35\pm 0.06\%$ ($Z=92$) coincides with
the values of the normalization correction $-4.5\pm 0.2\%$ for
$2\%L_{\scriptscriptstyle R}$ lead target and $-5.6\pm 0.3\%$ for
$3\%L_{\scriptscriptstyle R}$ uranium target, respectively, within
the experimental error.

\item
The sample average over the range $1\leq s \leq \infty$,
$\bar\delta_{\scriptscriptstyle CC}\left[\langle
dI/d\omega\rangle\right]=-4.70\pm 0.49\%$,
 excellent agrees in the regime of small LPM suppression
 with the mean normalization correction
 $-4.7\pm 2\%$ obtained for 25 GeV data in the experiment  \cite{C-1.1.2}.

\item
Thus, we managed to show that the discussed discrepancy between
theory and experiment can be explained on the basis of the obtained
Coulomb corrections to the Born bremsstrahlung rate within the
Migdal LPM effect theory.

\item

This means  that applying the revised multiple scattering theory by
Moli\`{e}re allows one to avoid multiplying theore\-tical results by
above normalization factor and leads to agreement between the Migdal
LPM effect theory and experimental data \cite{C-1.1.1,C-1.1.2} for
sufficiently thick high Z targets over the range $20<\omega<500$
MeV.

\item
We evaluated the accuracy of the Fokker--Planck approach and the
Gaussian first-order representation of the distribution function
$w_0(\boldsymbol{\vartheta})$ in the limiting case $\omega=0$, and
we showed the need of the second-order correction of order of
$1/B^{\scriptscriptstyle \mathrm{B}}\sim 12\%$ for
$w(\boldsymbol{\vartheta})$ to eliminate the discrepancy between the
theory and experimental data over the frequency range $5<\omega<30$
MeV for 25 GeV beam and $0.7\%L_{\scriptscriptstyle \mathrm{R}}$
gold target of the experiment \cite{C-1.1.1,C-1.1.2}.

\item
Finally, we found the numerical results for the relative corrections
$\delta_{\scriptscriptstyle
\mathrm{CC}}\left[(dI/d\omega)_0\right]$,
$\delta_{\scriptscriptstyle \mathrm{CC}}\left[\Phi(s) \right]$, and
$\delta_{\scriptscriptstyle \mathrm{CC}}\left[\langle
dI/d\omega\rangle \right]$ in the LPM effect theory analogue for a
thin target over the range $5<\omega<30$  and demonstrated that
these corrections additionally improve the agreement between the
theory \cite{Shul'ga,Fomin} and experiment \cite{C-1.1.1,C-1.1.2}.

\end{itemize}

 {\bf Acknowledgments.} This work is devoted to the memory of
Alexander Tarasov. We would like to thank Spencer Klein (Lawrence
Berkeley National Laboratory, USA) for valuable information and
useful comments about some aspects of the SLAC E-146 experiment. We
thank Bronislav Zakharov  for drawing our attention to his more
recent works \cite{Z2}. We are grateful to Alexander Dorokhov for
his attention to this work and reading the manuscript.

\end{document}